\documentclass[12pt]{article}
\usepackage{graphicx,amsmath,bbm,physics,mathtools,mathrsfs}
\usepackage[T1]{fontenc}
\usepackage{txfonts}
\usepackage[justification=centering,format=plain, singlelinecheck=false]{caption}
\usepackage{subcaption}
\usepackage{hyperref}
\usepackage{multirow}
\usepackage{array}
\newcommand*{\Perm}[2]{{}^{#1}\!P_{#2}}%
\newcommand*{\Comb}[2]{{}^{#1}C_{#2}}%

\begin{document}

	
	\title{Quantum Multiphoton Rabi Oscillations in Waveguide QED}
	\date{}
	

\begin{center}
   \Large Quantum Multiphoton Rabi Oscillations in Waveguide QED\\
\end{center}

\begin{center}
  Debsuvra Mukhopadhyay\footnote{debsuvra@wustl.edu} ~and Jung-Tsung Shen\footnote{jushen@wustl.edu },\\
  \vspace{4pt}
  \textit{Department of Electrical and Systems Engineering, 
      Washington University in St. Louis, St. Louis, Missouri 63130, USA.}
\end{center}

\pagenumbering{arabic}

\begin{abstract}
	The future of quantum information processing hinges on chip-scale nanophotonics, specifically cavity QED and waveguide QED. One of the foremost processes underpinning quantum photonic technologies is the phenomenon of Rabi oscillations, which manifests when a qubit is irradiated by an intense laser source. Departing from the conventional semiclassical framework, we expound on the more general, quantum-theoretic case where the optical excitation takes the form of a multiphoton Fock state, and the qubit couples to a continuum of radiation modes. By employing the real-space formalism, we analytically explore the scattering dynamics of the photonic Fock state as it interfaces with a two-level emitter. The resulting amplitude for atomic excitation features a linear superposition of various independent scattering events that are triggered by the potential of sequential photon absorptions and emissions. The lowest-order excitation event, initiated by the stochastic scattering of one of the several photons, aptly characterizes the dynamics in a weak-field environment. This event is complemented by a multitude of higher-order scattering events ensuing from repeated atom-photon interactions. The temporal evolution of the qubit excitation in our configuration closely mirrors the semiclassical predictions, particularly in the strong-pumping limit where Rabi oscillations unfold. Notably, this compatibility with the semiclassical paradigm transcends beyond the strong-excitation regime and applies both to the weak-driving and large-detuning limits. In a nutshell, our analysis extends the existing results on quantum Rabi oscillations pertinent to single-mode cavity QED, to the multimode, waveguide-QED configurations wherein flying photons are the information carriers.  Additionally, we delve into the dynamics of pulsed wave packets, shedding light on the potential to substantially enhance excitation efficiency, even in scenarios involving just a few photons. Beyond their theoretical merit, these findings should hold practical relevance for future Fock-state-based quantum computing and emerging waveguide-integrated photonic technologies such as those involving superconducting circuitry. 
	
\end{abstract}

\section{Introduction} \label{intro}
\begin{figure}
    \centering
    \includegraphics[scale=0.2]{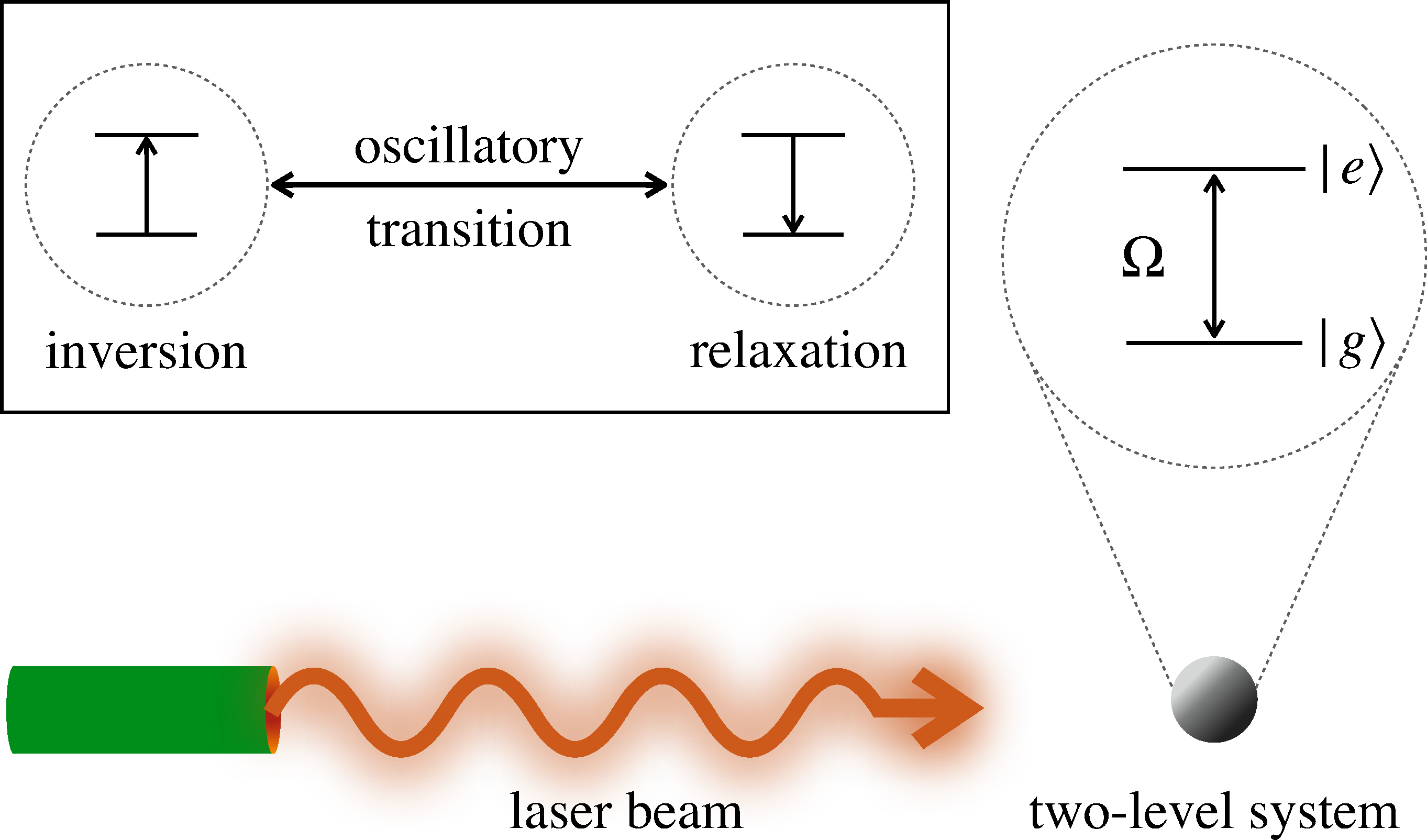}
  \caption{Semiclassical model redux: a collimated laser jet incident on a point-like atomic dipole. The inset shows the atom undergoing cyclic dynamics, oscillating between its ground and excited states upon alternatively absorbing and discharging energy.}
	\label{figsemiclassical}
\end{figure}

As one of the central planks of optics, the phenomenon of Rabi flopping in a two-level system (illustrated in Fig. \ref{figsemiclassical}) subjected to an intense light field has piqued the curiosity of physicists since long. Originally formulated as a prototypical yet supremely potent semiclassical problem \cite{rabi}, this was later extended to a quantum mechanical description for a qubit wedged inside a cavity resonator \cite{jc}. Some of the earliest experimental confirmations of this phenomenon involved the observation of self-induced transparency \cite{sit}, the oscillatory resonance fluorescence from Rubidium atoms \cite{reso1,reso2}, and the appearance of Mollow triplets \cite{mollow}. Since nuclear spins in an oscillating magnetic field exhibit the same characteristics, the coherent manipulation of molecular spins has emerged as a leading resource for quantum state preparation via nuclear magnetic resonance (NMR) spectroscopy. The photonic interaction with a discrete, sharply peaked, atomic transition line has been at the root of many seismic discoveries in nanophotonics and photonic technologies, including, but not limited to, photon  anti-bunching \cite{antibunch1,antibunch2}, interference effects due to photonic indistinguishability \cite{interference1, interference2}, as well as on-demand single-photon generation in a solid-state environment \cite{antibunch1,interference1, sp1, sp2}. Lately, the feasibility of generating multiphoton states \cite{mp1,mp2} has also been explored with regard to various applications.

While the predictions of the Jaynes-Cummings model in cavity QED are strikingly consistent with the original semiclassical one, the periodic boundary conditions in a resonator ensure that the atom is coupled to to a single-mode, stationary electromagnetic wave. This could be contrasted against the semiclassical Rabi problem which features the scattering of an itinerant electromagnetic wave off the atomic dipole. Viewed quantum mechanically, light scattering in free space would always entail extraneous dissipative effects such as spontaneous emission from the atom. This stems from the inherent fuzziness of the atomic transition line, and ergo, the atom can emit into a continuum of radiation modes proximal to the resonance frequency. When an excited atom decays spontaneously to a lower energy level, the atomic coherence is impaired and the Rabi oscillations get stamped out quickly. As a general principle, the coherent flopping dynamics would survive insofar as the interaction time remains subordinate to the relaxation lifetime. In order to establish deeper analogies with the free-space scattering paradigm apropos of the semiclassical treatment, it would be of immense scientific interest to develop a systematic, fully quantum mechanical prescription to analyze scattering problems in the multiphoton regime. Particularly intriguing would be to assess the \textit{semiclassical regime} of asymptotically large photon numbers wherein the Rabi oscillations ought to be recovered. Current approaches addressing multiphoton scattering in free space typically involve using the Heisenberg-Langevin formalism and the associated input-output relations to predict the scattered output \cite{roy}. However, their applicability is mostly limited to special kinds of light sources with very specific photon correlation properties, such as coherent states which are classical-like in nature.

In this work, we set forth a new formulation to theoretically investigate the scattering dynamics of \textit{arbitrary}, \textit{propagating}, single-mode, multiphoton excitations across an atomic dipole. The radiation sources could be Fock states featuring arbitrarily large but definite quanta of radiation, or any generic superposition of Fock states, inclusive of coherent states. Since the atomic diameter is typically much shorter than the optical wavelength, the isotropic s-wave scattering contributes predominantly to the scattering process, accordingly reducing the three-dimensional problem to an effectively one-dimensional setting. In light of that, we contextualize our analysis in a one-dimensional, chiral-waveguide configuration which supports the unidirectional propagation of free photons. As a quick and compact summary, we enumerate, in Table. \ref{table1}, some of the salient hallmarks of the semiclassical, the cavity-QED, and the waveguide-QED formalisms, which are of great relevance to the theoretical modeling of light-matter interaction. With the advent of photonic technologies enabling the controlled production of multiphoton Fock states, there has been an upswing of interest in the utility of photon-photon correlations naturally spawned by a scattering agent. Beyond their theoretical significance, these nonclassical photon-correlated states could act as advantageous resources for quantum simulations \cite{sturges}, quantum metrology \cite{metrology1,metrology2}, and imaging superresolution beyond the diffraction limit \cite{imaging1,imaging2}, and could even be tailored to prepare squeezed Schrödinger cat states \cite{catstates}.\\

\begin{table}[h!]
	\centering
	\begin{tabular}{ | m{12em} | m{12em}| m{12em} |  }
		\hline
	\centering	\vspace{2.5mm} Semiclassical Model & \centering \vspace{2.5mm} Cavity QED & \vspace{2.5mm} \hspace{10mm} Waveguide QED\\ [1.5ex]
		\hline\hline
		\small Light is treated classically; associated Hilbert space is two-dimensional. & \small Light is quantized; Hilbert space is infinite-dimensional with a discrete basis. & \small Light is quantized; Hilbert space is infinite-dimensional, spanned by a continuously distributed basis set.\\
		\hline
		\small Atom interacts with a collimated laser jet. & \small Atom couples to a single, discrete resonator mode. & \small Atom is submerged within a continuum of electromagnetic modes.\\
		\hline
		\small Interaction Hamiltonian goes as $H_{\text{int}}=\hbar g(\sigma_++\sigma_-)$  & \small $H_{\text{int}}=\hbar g_k(\sigma_+c_k+h.c.)$  &\small $H_{\text{int}}=\hbar \int\dd{k}g(k)\hspace{1mm}(\sigma_+c_k+h.c.)$\\
		\hline
		\small Free-space scattering problem;  source emits traveling waves. &  \small Periodic boundary condition results in stationary waves. & \small Open boundary conditions imply propagating photons.\\
		\hline
		\small Spontaneous emission cannot be modeled & \small Spontaneous emission can be incorporated extraneously. & \small Atomic relaxation is naturally accounted for.\\
		\hline
	\end{tabular}
	\caption{Comparative chart summarizing the key similarities and differences across the semiclassical, cavity-based, and waveguide-based paradigms for probing light-atom interaction.}
	\label{table1}
\end{table}

From a purely theoretical perspective, analyzing the multiphoton regime of waveguide QED requires highly nuanced considerations, which partly explains why advances in this direction have been far and few between. Not too long back, the exact eigenstates of a multiphoton waveguide-QED system and the mathematical structures thereof for arbitrarily large photon numbers were analytically deconstructed \cite{fockstate}. Very recently, a theoretical scheme to deterministically engineer traveling Fock states in a chiral configuration was conceptualized \cite{liu}. In this article, we present our detailed investigation on the dynamical response of a qubit due to an arbitrary multiphoton radiation line source. The wave equations ensuing from the time-dependent Schrödinger equation are highly nontrivial and heavily coupled; nonetheless, we can work out exact analytical solutions to these equations without appealing to any intermediate algebraic approximations. By employing an iterative procedure, we obtain the simplified expansion for the time-varying wave function that pertinently determines the excitation amplitude for the two-level system. The general methodology developed to this end facilitates solving the dynamics for any arbitrary initial wave packet. Generally speaking, the wave function consists of a linear superposition of various excitation orders which independently account for the different, mutually exclusive, scattering events. Since the assumed configuration is chiral, a photon once scattered off never re-interacts with the atom. Consequently, for an incident $N$-photon Fock state, the excitation orders collectively embody a total of $N$ possible scattering pathways or interfering channels. To exemplify, if the input state is the single-mode, three-photon Fock state $\ket{3}_k$, where $k$ denotes the mode frequency, the exhaustive set of excitation pathways would be as follows: (i) the atom absorbs one of the photons randomly; (ii) the atom scatters off one of the photons and absorbs a second one; and (iii) the atom scatters off two photons in succession whereupon it absorbs the third photon and settles into its excited state. The simplest scenario which proceeds purely via the absorption of a single photon (without any subsequent emission) directly corresponds to the weak-excitation regime and can be reconciled with the Fermi's Golden Rule. The higher-order excitation pathways become progressively more important with the increase in photon numbers. Specifically, in the strong pumping regime characterized by the limit $N\rightarrow\infty$, all of the scattering orders attain comparable significance and elegantly interfere to yield a sinusoidal excitation probability with the appropriate Rabi frequency. This excellent agreement with the semiclassical predictions serves as a validity check for the overarching formalism we have developed to analyze multiphoton scattering dynamics. Finally, in the few-photon regime, we establish the utility of leveraging pulsed light fields in starkly enhancing the atomic absorption probability in contrast to quasimonochromatic sources which hardly excite the atom. To summarize, therefore, our treatment differs considerably from more traditional approaches not only in that it lends a concise recipe for investigating scattering problems down to the few-photon sector, but also because it sets down the crossover between the quantum domain of finite photon numbers and the semiclassical regime of asymptotically large photon numbers on a firm theoretical foundation.

In Sec. \ref{S2},  we set forth our extensive formulation of the scattering problem by introducing the relevant configuration, and develop, in Sec. \ref{S3}, its dynamical solution for a stream of monochromatic, phase-coherent, photonic plane waves impinged on a two-level system. The physical relevance of the various interfering pathways as well as their asymptotic simplifications in some special limits are discussed. In Sec. \ref{S4}, we extend our dynamical analysis to the case of quasimonochromatic multiphoton wave packets, subsequent to which, we derive the probability of atomic excitation in Sec. \ref{S5}. The correct limiting behaviors of the semiclassical model are appropriately recovered. Lastly, in Sec. \ref{S6}, we briefly outline the scattering characteristics of pulsed, few-photon wave packets incident on an atomic dipole. The key results are summarized in Sec. \ref{S7}.

\section{Real-space formulation of multiphoton scattering in waveguide QED} \label{S2}
\begin{figure}
	\captionsetup{justification=raggedright,singlelinecheck=false}
	\centering
	\includegraphics[scale=0.2]{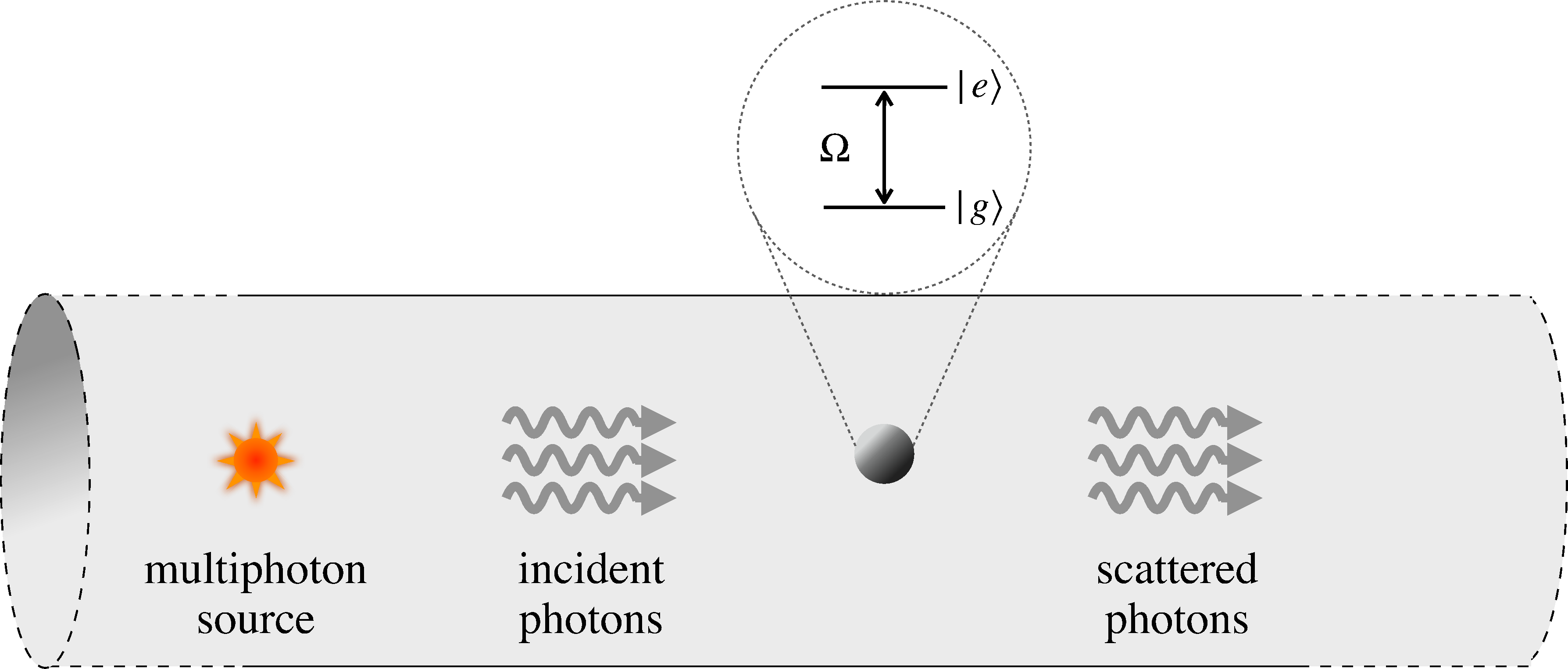}
	\caption{Schematic of a two-level atom embedded inside a transmission line or a waveguide. A multiphoton line source injects a flurry of photons into the waveguide channel which couple strongly to the atomic transition and are eventually scattered off.}
	\label{figwaveguidemodel}
\end{figure}

The setup considered in our work is portrayed in Fig. \ref{figwaveguidemodel}, which typifies a one-dimensional (1D) waveguide coupled to a two-level system. Such a two-level system can have several practical realizations, including, for example, a quantum dot \cite{qdot1,qdot2}, a superconducting qubit, a nitrogen vacancy center \cite{nvcenter}, or a cold atom \cite{atom1,atom2}. For conciseness, we categorize either of them as a ``two-level atom", whether natural or artificial. A 1D waveguide configuration could be similarly realized in a variety of architectures such as a microwave transmission line \cite{line}, a line defect in a photonic bandgap crystal \cite{crystal}, or an optical fiber. A quintessential figure of merit describing a robust waveguide-QED framework is the $\beta$ factor which concerns the efficiency of spontaneous emission into the waveguide. In the ideal scenario, featuring a sufficiently large $\beta$ factor usually means that lateral losses from the atom remain strongly suppressed in contrast to the guided emission. Furthermore, we also consider, for simplicity, a single-polarization single-mode (SPSM) waveguide \cite{SPSM} in order that unwanted photonic mode conversions can hardly impair quantum interference effects. 

The Hamiltonian of the composite waveguide-QED system can transformed into real space \cite{shen1,shen2} via Fourier transforming its $k$-space representation, leading to the form 
\begin{align}
	H=&-i\hbar v_g\int\dd{x}c_R^{\dagger}(x)\frac{\partial}{\partial x}c_R(x)+i\hbar v_g\int\dd{x}c_L^{\dagger}(x)\frac{\partial}{\partial x}c_L(x)+\hbar\Omega\ket{+}\bra{+}\hspace{1mm}\notag\\
	&+\hbar\int\dd{x}\overline{\mathcal{V}}\delta(x)\bigg[\bigg(c_R^{\dagger}(x)+c_L^{\dagger}(x)\bigg)\sigma_-+\bigg(c_R(x)+c_L(x)\bigg)\sigma_+\bigg], \label{e1}
\end{align}
where the rotating-wave approximation has been employed and consequently counter-rotating terms neglected. Here, $v_g>0$ is the group velocity of the propagating field, $c_R^{\dagger}(x)$ ($c_L^{\dagger}(x)$) represents the bosonic operators that create a right- (left-) moving photon at the position $x$, $\sigma_{\pm}$ denote the fermionic ladder operators that excite and deexcite the atom respectively, $\ket{+}$ is the atomic excited state with $\Omega$ equaling the atomic transition frequency, and $\overline{\mathcal{V}}$ signifies the strength of atom-photon interaction. The explicit polarization indices for the participating photons in the SPSM waveguide are witheld for convenience. It can be shown that $\Gamma=\overline{\mathcal{V}}^2/v_g$ yields the rate of spontaneous atomic relaxation into the waveguide modes and also characterizes the natural linewidth of the transmission spectrum \cite{shen1}. While this describes the more general nonchiral model wherein photons propagate freely in either direction, we can switch to a canonically equivalent chiral representation \cite{shen3} of the system by executing the following transformations:
\begin{align}
c_e(x)&=\frac{c_R(x)+c_L(-x)}{\sqrt{2}},\notag\\
c_o(x)&=\frac{c_R(x)-c_L(-x)}{\sqrt{2}}. \label{e2}
\end{align}
This enables the decomposition of the full Hamiltonian into two independent ``even" and ``odd" components,
\begin{align}
H_e&=-i\hbar v_g\int\dd{x}c_e^{\dagger}(x)\frac{\partial}{\partial x}c_e(x)+\hbar\Omega\ket{+}\bra{+}\hspace{1mm}+\hbar\int\dd{x}\mathcal{V}\delta(x)[c_e^{\dagger}(x)\sigma_-+c_e(x)\sigma_+],\label{e3}\\
H_o&=-i\hbar v_g\int\dd{x}c_o^{\dagger}(x)\frac{\partial}{\partial x}c_o(x), \label{e4}
\end{align}
where $\mathcal{V}=\sqrt{2}\hspace{0.8mm}\overline{\mathcal{V}}$ denotes the coupling strength between the atom and the even modes. Further, the Hamiltonians $H_e$ and $H_o$ commute, implying that the ensuing unitary evolution operator $\mathcal{U}(t)=e^{-iHt/\hbar}$ can be written as a simple tensor product of two local unitary operators in the even and the odd subspaces respectively,
\begin{align}
\mathcal{U}(t)=\mathcal{U}_e(t)\otimes\mathcal{U}_o(t) \label{e5}.
\end{align}
Notably, the odd component is also decoupled from the atomic system, implying free evolution in the odd subspace. The coherent exchange of energy between the atom and the photons is thereby limited to the even subspace, which effectively underpins a chiral model with unidirectional photon propagation \footnote{It is worth remarking here that decomposing the original field operators is akin to superposing left- and right-propagating sinusoidal waves of the form $e^{ikx}$ and $e^{-ikx}$. The combined wave with even parity, $e^{ikx}+e^{-ikx}\sim\cos x$ does interact with the atom located at $x=0$, while the combination with odd parity, $e^{ikx}-e^{-ikx}\sim\sin x$, has a node at the atomic location.}. From a mathematical perspective, obtaining a complete set of solutions in both the even and the odd subspaces permits solving the scattering problem in the original nonchiral system \cite{fockstate,shen3}. Fortuitously, there also exist physical systems which can be precisely described by a chiral model. For instance, by engineering photonic analogs of the quantum Hall effect \cite{chiral1,chiral2} in photonic crystals, the backscattered modes can be topologically suppressed producing chiral edge channels. Alternatively, the property of spin-momentum locking in evanescently coupled photons can be harnessed in the efficient guiding of directional spontaneous emission along a waveguide, as has been demonstrated particularly in dielectric nanostructures \cite{chiral3,chiral4,chiral5}. Either of these scenarios can be exactly treated within the fully chiral framework. In view of these realistic possibilities, we, henceforth, restrict our investigations to the fully chiral model as described by Eq. \ref{e3}. So we rewrite the system Hamiltonian by suppressing the subscript $e$ in Eq. \ref{e3}:
\begin{align}
	\mathcal{H}=-i\hbar v_g\int\dd{x}c^{\dagger}(x)\frac{\partial}{\partial x}c(x)+\hbar\Omega\ket{+}\bra{+}\hspace{1mm}+\hbar\int\dd{x}\mathcal{V}\delta(x)[c^{\dagger}(x)\sigma_-+c(x)\sigma_+]. \label{e6}
\end{align}
In the absence of backscattering, any photon, following its absorption by the atom, can be emitted only in the forward direction into a narrow band of the radiation continuum centered around the atomic transition frequency $\Omega$. The above Hamiltonian is excitation-preserving since the total excitation operator $\mathcal{N}=\int\dd{x}c^{\dagger}(x)c(x)+\hspace{1mm}\bigg(\ket{+}\bra{+}\hspace{1mm}+\hspace{1mm}\ket{-}\bra{-}\bigg)\hspace{1mm}$ commutes with the system Hamiltonian, i.e., $[\mathcal{N},\mathcal{H}]=0$. Therefore, if the atom is excited by a fixed quanta of the radiation field, any generic, time-dependent quantum state would be expressible in the form
\begin{align}
	\ket{\psi_N(t)}=\int\dd{\va{x}_{N}}f(\va{x}_N;t)\ket{\va{x}_N;-}+\int\dd{\va{x}_{N-1}}e(\va{x}_{N-1};t)e^{-i\Omega t}\ket{\va{x}_{N-1};+}, \label{e7}
\end{align}
where $N$ equals the input photon number, while $f(\va{x}_N;t)$ and $e(\va{x}_{N-1};t)$ determine the dynamically evolving wave functions of the $N$-photon and ($N-1$)-photon components of the composite state. For brevity, the pithy algebraic notations $\va{x}_n=(x_1,x_2,x_3,\hdots,x_n)$ and $\dd{\va{x}_n}=\prod_{i=1}^n\dd{x}_i$ have been used. Here $\ket{\va{x}_n}=\dfrac{1}{\sqrt{n!}}\bigg(\prod_{j=1}^nc^{\dagger}(x_j)\bigg)\ket{0}$ represents an $n$-photon quantum state written in the second-quantized representation with photonic occupation at discrete spatial locations, with $\ket{0}$ identifying the electromagnetic vacuum. For a quick reference, some of the elementary mathematical properties of these states are listed in Appendix A. For mathematical convenience, the coefficient $e^{-i\Omega t}$ encoding the effect of free evolution on the atomic excited state has been factored out from $e(\va{x}_{N-1};t)$ in the second term. 

The time-dependent Schr$\ddot{\text{o}}$dinger equation $i\hbar\dfrac{\dd}{\dd t}\ket{\psi_N(t)}=H\ket{\psi_N(t)}$, underlying the system dynamics, can be simplified by taking projections along the two basis vectors $\ket{\va{x}_N;-}$ and $\ket{\va{x}_{N-1};+}$, which precipitate in a set of coupled wave equations,
\begin{align}
	\bigg(\frac{\partial}{\partial t}+v_g\sum_{j=1}^N\frac{\partial}{\partial x_j}\bigg)f(\va{x}_N;t)&=-\frac{i\mathcal{V}e^{-i\Omega t}}{\sqrt{N}}\sum_{j=1}^N\delta(x_j)\hspace{1mm}e(\{x_i\}\setminus x_j; t),\label{e8}\\
	\bigg(\frac{\partial}{\partial t}+v_g\sum_{j=1}^{N-1}\frac{\partial}{\partial x_j}+\Gamma\bigg)e(\va{x}_{N-1};t)&=-i\sqrt{N}\mathcal{V}e^{i\Omega t}f(x_1,x_2,...x_{N-1},0^-;t), \label{e9}
\end{align} 
wherein the effective coupling parameter $\Gamma=\mathcal{V}^2/(2v_g)$ has naturally cropped up. Note that in the set-theoretic parlance, one can define a set $Y=A\setminus B$ that contains all the elements of $A$ except for those belonging to $B$. Thus, the sum $\sum_{j=1}^N\delta(x_j)\hspace{1mm}e(\{x_i\}\setminus x_j; t)$ expands out into the symmetrized expression
\begin{align*}
	\delta(x_1)e(x_2,x_3,...,x_N;t)\hspace{1mm}+\hspace{1mm}\delta(x_2)e(x_1,x_3,...,x_N;t)\hspace{1mm}+\hspace{1mm}\hdots\hspace{1mm}\delta(x_N)e(x_1,x_2,...,x_{N-1};t).
\end{align*}
In deriving the coupled wave equations, the permutation symmetry apropos of bosonic statistics, supplemented by the key boundary condition,
\begin{align}
	f(\va{x}_{N-1},0^+;t)=f(\va{x}_{N-1},0^-;t)-\dfrac{i\mathcal{V}e^{-i\Omega t}}{\sqrt{N}v_g}e(\va{x}_{N-1};t),\label{e10}
\end{align}
has been invoked. Such a jump discontinuity in $f(\va{x}_N;t)$ across any two adjoining hyperoctants in the $N$-dimensional hyperspace follows from the $\delta$-type local interaction term. It can be derived forthwith from the first differential equation, i.e., Eq. \ref{e8}, by integrating both sides over an infinitesimally small range from $x_N=-\varepsilon$ to $x_N=+\varepsilon$, for $\varepsilon\rightarrow 0$. As a general principle that we shall adopt in our analysis, we can solve for the amplitude $e(\va{x}_{N-1};t)$ by eliminating $f$-dependent terms from Eqs. \ref{e8}, \ref{e9}. This approach leads to a self-consistent equation involving the desired wave function which can then be worked out subject to appropriate initial conditions.

\section{Scattering dynamics: Plane waves} \label{S3}

\begin{figure}
	\captionsetup{justification=raggedright,singlelinecheck=false}
	\centering
	\includegraphics[scale=0.2]{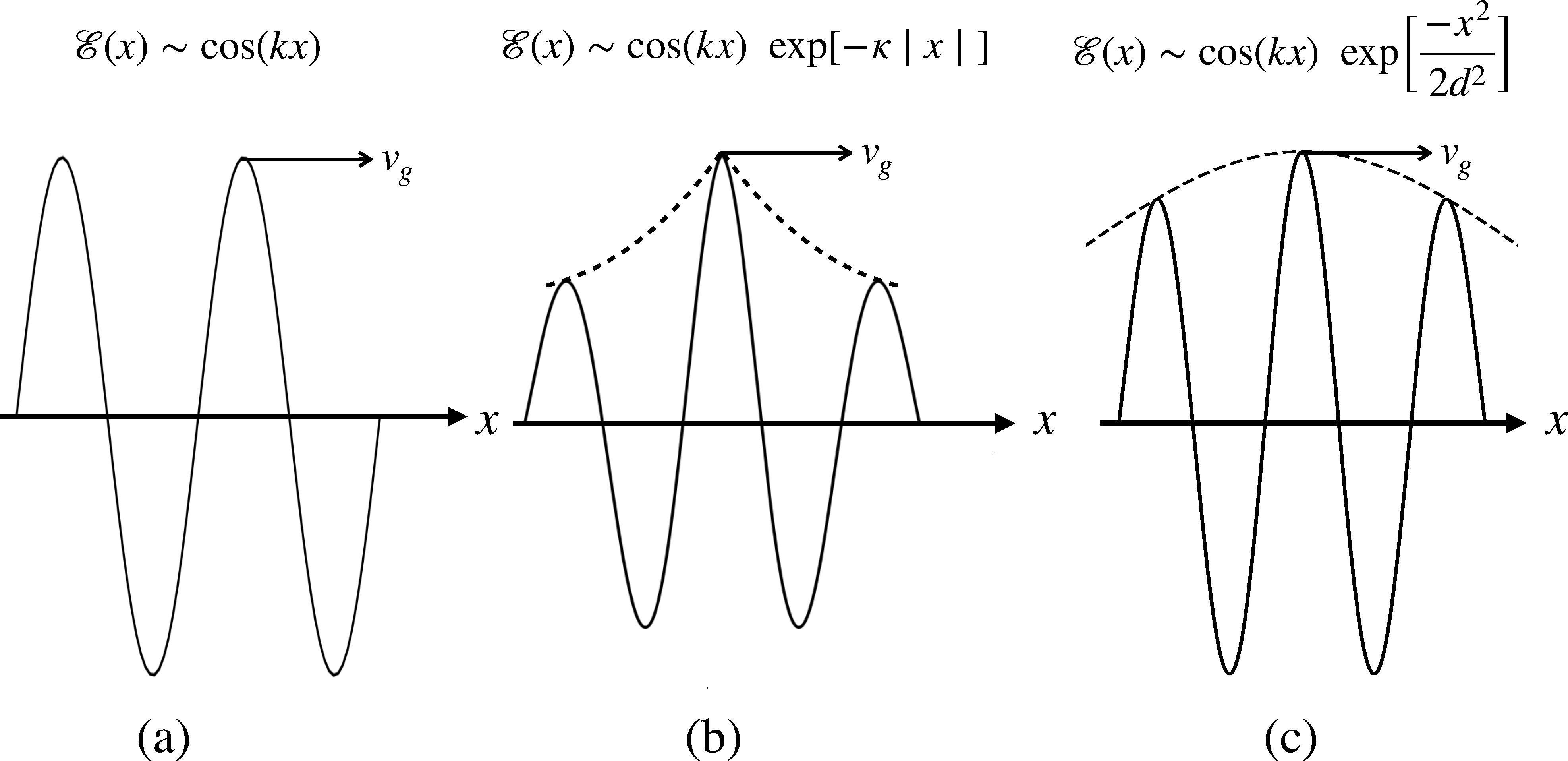}
	\caption{Real-space representations of some typical electromagnetic fields: (a) plane wave, (b) Lorentzian wave packet, and (c) Gaussian wave packet. A Lorentzian packet in $k$-space gets mapped into a damped plane wave in the position space. The parameter $\kappa$ signifies the damping rate of this wave, while $d$ represents the width of the Gaussian envelope. Either of these waves moves rightward with a speed $v_g$ until it encounters a scattering potential. We discuss the solution to plane-wave scattering in Sec. \ref{S3} while the two other cases will be addressed in Sec. \ref{S4}.}
	\label{figplane}
\end{figure}
In order to address the quantum scattering problem, we focus on the canonical example of a single-mode, multiphoton Fock state which begins to scatter off the two-level atom at $t=0$. Since Fock states span the entire photonic Hilbert space, this computation enables us to solve the dynamics due to any incident radiation state, including coherent states of light which possess well-defined amplitudes and phases. Now single-mode Fock states have a plane-wave representation in position space, as can be elucidated below (Fig. \ref{figplane} depicts three most important types of waves). If the atom starts out in its ground energy level, the initial quantum state of the joint system takes the form
\begin{align}
	\ket{\psi_N(0)}=\ket{N}_k\otimes\ket{-}=\frac{1}{\sqrt{N!}}(c_k^{\dagger})^N\ket{0;-}=\frac{1}{(2\pi)^{N/2}}\int\dd{\va{x}_{N}}e^{ik\sum_{j=1}^Nx_j}\ket{\va{x}_N;-}, \label{e11}
\end{align}
where $k$ is the wavenumber of the incident field, and $c_k^{\dagger}=\dfrac{1}{\sqrt{2\pi}}\int_{-\infty}^{\infty}\dd{x}e^{ikx}c^{\dagger}(x)$ is the frequency-space photon creation operator. Thence, it follows that the initial conditions on the wave functions $f(\va{x}_N;0)=\bra{\va{x}_N;-}\ket{\psi_N(0)}$ and $e(\va{x}_{N-1};0)=\bra{\va{x}_{N-1};+}\ket{\psi_N(0)}$ can be embodied as
\begin{align}
	f(\va{x}_N;0)&=\frac{1}{(2\pi)^{N/2}}e^{ik\sum_{j=1}^Nx_j},\notag\\
	e(\va{x}_{N-1};0)&=0. \label{e12}
\end{align}
Armed with specific initial conditions, we now proceed to solve the wave equations. Eq. \ref{e8} can be formally integrated out by using the method of Fourier transform to obtain $f(\va{x}_N;t)$ in terms of two contributing sources, (i) the evolving waveform of the free field, and (ii) a term underscoring the superposition of the wave function $e$ evaluated at \textit{retarded} spacetime points, which encodes the effect of interaction (c.f. Appendix B):
\begin{align}
	f(\va{x}_{N};t)&=f\bigg(\{x_i-v_gt\};0\bigg)-\frac{i\mathcal{V}e^{-i\Omega t}}{\sqrt{N}v_g}\sum_{j=1}^Ne^{i(\Omega/v_g)x_j}e\bigg(\{x_i-x_j\}_{i\neq j};t-\frac{x_j}{v_g}\bigg)\Theta(x_j)\Theta\bigg(t-\frac{x_j}{v_g}\bigg). \label{e13}
\end{align}
Here, $f(\{x_i-v_gt\};0)=f(x_1-v_gt,x_2-v_gt,\hdots,x_N-v_gt)$ represents the freely-evolving wave function that solves the source-free homogeneous wave equation, and $\Theta(.)$ stands for the Heaviside-theta function. As an example, the term corresponding to $j=3$ in the above summation, would go as $e^{i(\Omega/v_g)x_3}e(x_1-x_3,x_2-x_3,x_4-x_3,...,x_N-x_3; t-{x_3}/{v_g})\Theta(x_3)\Theta(t-{x_3}/{v_g})$. Note that Eq. \ref{e13} implies that for all the spatial coordinates being negative, we have the trivial solution
\begin{align}
	f(\va{x}_{N};t)\bigg|_{x_i<0, \forall i}=f\bigg(\{x_i-v_gt\};0\bigg)=\frac{1}{(2\pi)^{N/2}}\prod_{j=1}^N e^{i(kx_j-\omega_kt)}, \label{e14}
\end{align}
in which $\omega_k=v_gk$ is the common frequency of incident photons. Likewise, Eq. \ref{e9} yields the formal solution
\begin{align}
	e(\va{x}_{N-1};t)=\underbrace{e\bigg(\{x_i-v_gt\};0\bigg)}_{=0}-i\sqrt{N}\mathcal{V}e^{i{\Omega} t}\int_{0}^{t}\dd\tau e^{-i\tilde{\Omega}\tau}f(\{x_i-v_g\tau\},0^-;t-\tau), \label{e15}
\end{align}
where $\tilde{\Omega}=\Omega-i\Gamma$, and $f(\{x_i-v_g\tau\},0^-; \hspace{1mm}t-\tau)$ is the condensed notation for $f(x_1-v_g\tau,x_2-v_g,\hdots,x_{N-1}-v_g\tau, \hspace{1mm}0^-;t-\tau)$. The first term drops out due to the initial condition on $e$, as stipulated by Eq. \ref{e12}. By substituting for the formal expression of $f$ from Eq. \ref{e13} into Eq. \ref{e15}, one finds a self-consistent equation involving purely the $e$-functions,
\begin{align}
	e(\va{x}_{N-1};t)=e_0(\va{x}_{N-1};t)-\frac{\mathcal{V}^2}{v_g}\sum_{j=1}^{N-1}e^{i(\Omega/v_g)x_j}\mathcal{T}_j(\va{x}_{N-1};t)\Theta(x_j)\Theta\bigg(t-\frac{x_j}{v_g}\bigg), \label{e16}
\end{align}
where we have expediently introduced two intermediate functions 
\begin{align}
	\mathcal{T}_j(\va{x}_{N-1};t)&=\int_0^{x_j/v_g}\dd\tau e^{-i\tilde{\Omega}\tau}e\bigg(x_1-x_j,x_2-x_j,...,x_{j-1}-x_j,\underbrace{v_g\tau-x_j}_{j^{th} location},x_{j+1}-x_j,...,x_{N-1}-x_j;t-\frac{x_j}{v_g}\bigg),\notag\\
	e_0(\va{x}_{N-1};t)&=\sqrt{N}\mathcal{V}e^{i\Omega t}\int_{0}^{t}\dd\tau e^{-i\tilde{\Omega}\tau}f(\{x_i-v_gt\},-v_g(t-\tau);0)\notag\\
	&=-\frac{\sqrt{N}i\mathcal{V}}{(2\pi)^{N/2}(\omega_k-\tilde{\Omega})}\bigg(\sum_{j=1}^{N-1}e^{ikx_j}\bigg)e^{-i(N\omega_k-\Omega)t}\bigg(e^{i(\omega_k-\tilde{\Omega})t}-1\bigg). \label{e17}
\end{align}
We remark that $e_0$ contains purely the signature of an effective ``single-photon" absorption that elevates the atom to its excited state. It is only through the more nontrivial set of functions $\{\mathcal{T}_j(\va{x}_{N-1};t)\}$ that any concrete information on cascaded photon absorption/emission events (as more and more photons get scattered off) gets encoded. In order to impart a better sense of the precise functionality of these $\mathcal{T}_j$ functions, we need to introduce some mathematical preliminaries which will be useful in finding the complete expansion of the function $e(\va{x}_{N-1};t)$. Since $e_0$ is the only known function in Eq. \ref{e16}, it is intuitively obvious that a closed-form expression for $e(\va{x}_{N-1};t)$ could, in principle, be built up purely in terms of this known function. To that end, we forge a mathematical dictionary that can help accomplish this objective. For the detailed development, we refer the reader to Appendix C and proceed directly to the solution in the forthcoming discussion.

\subsection{Analytical solution to the equations of motion}

\begin{table}[h!]
	\centering
	\begin{tabular}{ | m{10em} | m{6em}| m{6em} |  m{6em} |  m{6em} | }
		\hline
		\centering \vspace{1mm}Excitation orders\vspace{1mm} \hspace{2mm}$\rightarrow$ & \centering \vspace{1mm}Zeroth Order\vspace{1mm} & \centering \vspace{1mm}First order\vspace{1mm} & \centering \vspace{1mm}Second order\vspace{1mm} & \vspace{1mm}\hspace{3.5mm}Third order\vspace{1mm}\\ [0.5ex]
		\hline\hline
		\small \centering \vspace{1mm}Photons absorbed\vspace{1mm} & \small \centering \vspace{1mm}1\vspace{1mm}   & \small \centering \vspace{1mm}1\vspace{1mm} & \small \centering \vspace{1mm}1\vspace{1mm} & \small \vspace{1mm}\hspace{11.5mm}1\vspace{1mm}\\
		\hline
		\small \centering \vspace{1mm}Photons scattered away \vspace{1mm}& \small \centering \vspace{1mm}$0$\vspace{1mm} & \small \centering \vspace{1mm}$1$\vspace{1mm} & \small \centering \vspace{1mm}$2$\vspace{1mm} & \small \vspace{1mm}\hspace{11.5mm}$3$\vspace{1mm}\\
		\hline
		\small \centering \vspace{1mm}Non-participating photons\vspace{1mm} & \small \centering \vspace{1mm}$3$\vspace{1mm} & \small \centering \vspace{1mm}$2$\vspace{1mm} &\small \centering\vspace{1mm} $1$\vspace{1mm} & \vspace{1mm}\hspace{11.5mm}\small $0$\vspace{1mm}\\
		\hline
	\end{tabular}
	\caption{Photon scattering channels of various orders leading to atomic transition to the excited level, subject to the incidence of 4 identical, freely propagating photons. In the most general scenario, for an initial Fock state $\ket{N}_k$, the $n^{\text{th}}$ excitation order ($n<N$) relevant to the scattering mechanism would correspond to the sequential scattering (absorption followed by emission) of $n$ of these photons, with $1$ of the remaining photons absorbed and $N-n-1$ photons interacting with a saturated atom.}\label{table2}
	
\end{table}

Before we present a concrete analytical expression, we would like to delineate an intuitive picture of the excitation mechanism. For specificity, let us consider the four-photon scattering dynamics.  In principle, the atom, under these conditions, could transit to its excited level through four independent scattering mechanisms, as identified in Table \ref{table2}. The simplest (zeroth-order) pathway concerns the direct excitation by one of the photons. Any of the higher order pathways would involve a sequence of elementary absorption and emission mechanisms. For instance, the first-order excitation would unfold when the atom initially absorbed a photon but chose to emit it and subsequently absorbed a second photon to reach its excited state. The photon thus emitted could be labeled as a scattered photon. One can thereby anticipate that the time-dependent wave function $e(\va{x}_{N-1};t)$, which has the information regarding atomic excitation encoded into it, should be expressible as a linear superposition of individual wave functions each representing a distinct scattering channel. 

With this heuristic impression in mind, we can now proceed with an exact analytical treatment of Eq. \ref{e16}, subject to the definitions in Eq. \ref{e17}. The derivation, however, is veritably tedious, and thence, we directly present the analytical solution here. The computation is based on an algebraic formalism which is discussed at great length in Appendix C, subsequent to which, an iterative solution is worked out in Appendix D. This elaborate algebraic scheme helps to identify all of the higher-order excitation terms pertaining to Fock-state incidence. Notably, no intermediate approximations are incumbent in this derivation and the exact analytical expression, for $t>0$, is obtained to be
\begin{align}
	e(\va{x}_{N-1};t)=\mathcal{A}_0e^{-i(\omega_k-\Omega)t}&\prod_{l=1}^{N-1}e^{i(kx_l-\omega_kt)}\Bigg[\phi(t)+\sum_{j=1}^{N-1}\bigg(\frac{2i\Gamma}{\omega_k-\Omega+i\Gamma}\bigg)^j\sum_{\mathbf{\Lambda}_j\in\mathcal{P}_j}\phi(t^{(j)})\Theta(t^{(j)})\prod_{m=0}^{j-1}\phi\bigg(\frac{x_{\lambda_m}^{(m)}}{v_g}\bigg)\Theta(x_{\lambda_m}^{(m)})\hspace{1mm}\Bigg]\hspace{1mm}. \label{e18}
\end{align}
The expression above involves a number of new variables defined as follows,
\begin{align*}
	\mathcal{A}_0&=\frac{-\sqrt{N}i\mathcal{V}}{(2\pi)^{N/2}(\omega_k-\Omega+i\Gamma)},\\
	t^{(j)}&=t-x_{\lambda_{j-1}}/v_g,\\
	x_{\lambda_m}^{(m)}&=x_{\lambda_m}-x_{\lambda_{m-1}},\\
	\phi(t)&=e^{i(\omega_k-\Omega)t}e^{-\Gamma t}-1,
\end{align*}
with the boundary value $x_{\lambda_0}^{(0)}=x_{\lambda_0}$. Additionally, $\mathcal{P}_j$ is defined as the set of all possible permutations of any $j$ integers chosen from $\{1,2,3,...,N-1\}$, while $\mathbf{\Lambda}_j=(\lambda_0,\lambda_1,\lambda_2,\hdots,\lambda_{j-1})$ is a ($j-1$)-dimensional vector with integer elements. All of these variables are either introduced or derived systematically in the appendices C and D, and we refer the reader to the supplemental material thereof for a complete derivation of Eq. \ref{e18}. 

An alternative but rather appealing way to recast the above expression is by invoking the single-photon transmission coefficient for chiral propagation, $\mathbf{t}_k=\dfrac{\omega_k-\Omega-i\Gamma}{\omega_k-\Omega+i\Gamma}$. In terms of this transmission coefficient, we can rewrite Eq. \ref{e18} as
\begin{align}
			e(\va{x}_{N-1};t)=\mathcal{A}_0e^{-i(\omega_k-\Omega)t}&\prod_{l=1}^{N-1}e^{i(kx_l-\omega_kt)}\Bigg[\phi(t)+\sum_{j=1}^{N-1}(1-\mathbf{t}_k)^j\sum_{\mathbf{\Lambda}_j\in\mathcal{P}_j}\phi(t^{(j)})\Theta(t^{(j)})\prod_{m=0}^{j-1}\phi\bigg(\frac{x_{\lambda_m}^{(m)}}{v_g}\bigg)\Theta(x_{\lambda_m}^{(m)})\hspace{1mm}\Bigg]\hspace{1mm}.\label{e19}
		\end{align} 

\subsection{Composition of the wave function: Interference between multiple excitation channels}

We have already presented an intuitive tabular representation of the excitation mechanism in Table \ref{table2}, but now we have a full-fledged expression for the wave function consolidating our intuition. It is, therefore, worth characterizing and interpreting each of the constituent terms belonging to the full wave function in terms of the different excitation orders depicted earlier. To start off, we note that the wave function $e(\va{x}_{N-1};t)$ in Eq. \ref{e19} simplifies into some special functional forms in the various hyperoctants of the coordinate system spanned by the position variables $x_1,x_2,x_3,\hdots, x_{N-1}$. On account of the coordinate exchange symmetry, the wave function $e(\va{x}_{N-1};t)$ reduces to $N$ distinct functional forms in $N$ different spatial sectors that are identified by the number of positive coordinates in the set $\{x_1,x_2,x_3,\hdots, x_{N-1}\}$. In these sectors, we can define the specific forms to the wave function as
\begin{align}
    e(\va{x}_{N-1};t)=e^{(l)}(\va{x}_{N-1};t) ,\hspace{2mm} \text{if } n=l, \label{e20}
\end{align}
where ${n}\in\{0,1,2,\hdots,N-1\}$ denotes the number of positive-valued spatial coordinates in the photonic wave function, and the superscript $l$ labels the wave function in the subspace with exactly $l$ positive coordinates. For example, if $x_1$, $x_2$, and $x_3$ are positive in magnitude, while the rest of the coordinates $\{x_4,x_5,x_6,\hdots,x_{N-1}\}$ are all negative-valued, then we can use the function $e^{(3)}(x_1,x_2,x_3,\hdots,x_{N-1};t)$ to represent the appropriate wave function. Since the wave function respects permutation symmetry and there are $\Comb{N-1}{l}$ distinct combinations of having any $l$ coordinates positive, $e^{(l)}(\va{x}_{N-1};t)$ would accurately describe the functional representation in $\Comb{N-1}{l}$ out of the $2^{N-1}$ hyperoctants. A higher number of positive-valued photonic coordinates in the wave function would imply that higher-order interaction pathways are of relevance to the scattering mechanism. Intuitively, this makes perfect sense as any photon crossing the atomic interface has a non-zero probability of being absorbed the atom before it can be emitted forward via atomic relaxation. To draw a correspondence with our analytical expression, we can define the $l^{\text{th}}$-order contribution to the wave function as $K^{(l)}(\va{x}_{N-1};t)$, wherein
\begin{align}
	K^{(0)}(\va{x}_{N-1};t)&=e^{(0)}(\va{x}_{N-1};t)=\mathcal{A}_0e^{-i(\omega_k-\Omega)t}\prod_{q=1}^{N-1}e^{i(kx_q-\omega_kt)}\hspace{1mm}\phi(t),\notag\\
	K^{(l)}(\va{x}_{N-1};t)&=\mathcal{A}_0e^{-i(\omega_k-\Omega)t}\prod_{q=1}^{N-1}e^{i(kx_q-\omega_kt)}\hspace{1mm}(1-\mathbf{t}_k)^l\sum_{\mathbf{\Lambda}_l\in\mathcal{P}_l}\phi(t^{(l)})\Theta(t^{(l)})\prod_{m=0}^{l-1}\phi\bigg(\frac{x_{\lambda_m}^{(m)}}{v_g}\bigg)\Theta(x_{\lambda_m}^{(m)}),\hspace{3mm} \text{if } l\geq 1. \label{e21}
\end{align}
Then, the full wave function can be expanded out in terms of these indexed functions as
\begin{align}
	e(\va{x}_{N-1};t)=\sum_{l=0}^{N-1}K^{(l)}(\va{x}_{N-1};t), \label{e22}
\end{align}
 implying the stitching condition
 \begin{align}
 	e^{(l)}(\va{x}_{N-1};t)=e^{(l-1)}(\va{x}_{N-1};t)+K^{(l)}(\va{x}_{N-1};t). \label{e23}
 \end{align}
 
\begin{figure}
	\captionsetup{justification=raggedright,singlelinecheck=false}
	\centering
	\includegraphics[scale=0.25]{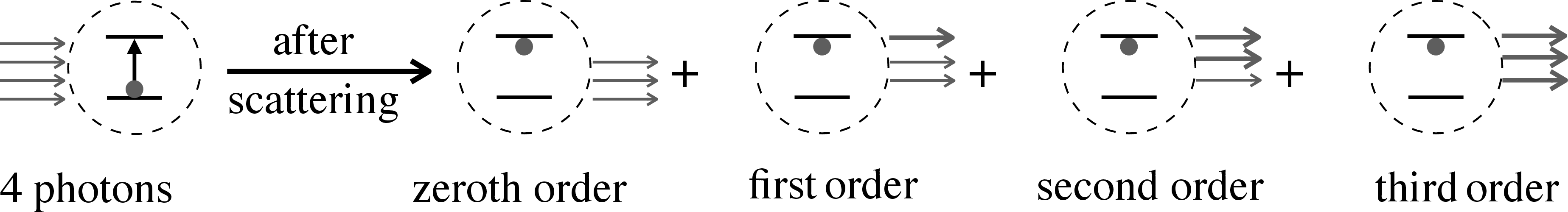}
	\caption{A pictorial representation of the four distinct scattering processes eliciting atomic transition to the excited state, for an initial four-photon Fock state incident on the atom. The number of scattered photons, i.e., photons that were previously absorbed and subsequently discharged, designates the excitation order. Thinner arrows indicate unabsorbed photons while thicker arrows signify scattered photons. 
	\vspace{5mm}}
	\label{figscattering}
\end{figure}

\begin{table}[h!]
	\centering
	\begin{tabular}{ | m{10em} | m{6em}| m{6em} |  m{6em} |  m{6em} | }
		\hline
		 & \centering \vspace{1mm} Zeroth Order \vspace{1mm} & \centering \vspace{1mm}First order\vspace{1mm} & \centering \vspace{1mm}Second order \vspace{1mm}& \vspace{1mm}\hspace{3.5mm}Third order\vspace{1mm}\\ [0.5ex]
		\hline\hline
		\small \vspace{1mm} \centering Photons scattered \vspace{1mm} & \small \centering \vspace{1mm} 0\vspace{1mm}   & \small \centering \vspace{1mm}1\vspace{1mm} & \small \centering \vspace{1mm}2\vspace{1mm} & \small  \vspace{1mm}\hspace{9.5mm} 3\\
		\hline
		\small \vspace{1mm}\centering Multiplicity of the scattering channel\vspace{1mm} & \small \centering \vspace{1mm}$1=\Perm{3}{0}$\vspace{1mm} & \small \centering \vspace{1mm}$3=\Perm{3}{1}$\vspace{1mm} & \small \centering\vspace{1mm} $6=\Perm{3}{2}$\vspace{1mm} & \small \vspace{1mm}\hspace{5.5mm} $6=\Perm{3}{3}$\vspace{1mm}\\
		\hline
		\small \vspace{1mm} \centering Net contribution to the wave function\vspace{1mm} & \small \centering \vspace{1mm}$K^{(0)}(x_1,x_2,x_3;t)$\vspace{1mm} & \small \centering \vspace{1mm}$K^{(1)}(x_1,x_2,x_3;t)$\vspace{1mm} &\small \centering \vspace{1mm}$K^{(2)}(x_1,x_2,x_3;t)$\vspace{1mm} & \small \vspace{1mm}$K^{(3)}(x_1,x_2,x_3;t)$\vspace{1mm}\\
		\hline
	\end{tabular}
	\vspace{1mm}
	\caption{Chart encompassing some basic properties of the distinct excitation pathways underlying the scattering dynamics of the incident Fock state $\ket{4}_k$. The total scattering wave function, which directly underpins the probability of atomic excitation, can be obtained as a linear superposition of four terms: $e(\va{x}_{3};t)=K^{(0)}(x_1,x_2,x_3;t)+K^{(1)}(x_1,x_2,x_3;t)+K^{(2)}(x_1,x_2,x_3;t)+K^{(3)}(x_1,x_2,x_3;t)$, where the function $K^{(l)}$ is defined in Eq. (22). The multiplicity of any of these scattering channels is determined by the number of ordered sequences in which the photons can be absorbed and subsequently emitted.\\}
	\label{table3}
\end{table}

To complement this discussion with a concrete example, we illustrate, in Fig. \ref{figscattering}, the contributions from the four distinct excitation pathways corresponding to the four-photon scattering dynamics, and lay out with the associated properties in Table \ref{table3}. Physically, the $l^{\text{th}}$-order interaction term underpins a component of the full scattering mechanism which describes a sequence of $l$ consecutive single-photon absorption-cum-emission events, followed by a final absorption event leading to the eventual excited state. This is engendered by the fact that the atom, upon having absorbed a photon and transited to the excited level, can emit it back into the radiation field, albeit in the forward direction owing to the chiral, unidirectional coupling. As an upshot of this, the emitted photon is barred from interacting any further with the atom. However, the atom is now free to absorb energy from any of the residual, hitherto unabsorbed photons in the field. Since the relaxation lifetime goes as $\Gamma^{-1}$, the atom would typically absorb successive photons at an interval comparable to its excited state lifetime. When a sequence of photons is scattered off in this fashion, the fermionic degree of freedom in the atom induces strong photon-photon correlations between the scattered photons while the remainder of the unabsorbed photons remain independent. This fundamentally changes the nature of the wave function. Thus the scattering amplitude $K^{(l)}(\va{x}_{N-1};t)$ is symbolic of the $l^{\text{th}}$-order excitation pathway representing $l$ number of scattered photons which propagate as a multipartite correlated system. Out of the remaining $N-l$ photons, one of the photons is annihilated upon absorption while the rest merely interface with a saturated atom. Furthermore, since the atom cannot re-interact with a previously emitted photon, it can, in principle, scatter up to $N-1$ photons into the right-moving channel before absorbing the final photon and settling into the excited level. This explains why the expansion of $e(\va{x}_{N-1};t)$, as derived in Appendix D, had to terminate after a finite number of iterations, in order for it to make physical sense.

The underlying physics of the scattering process lends itself to visually intuitive interpretations in two widely disparate regimes. When the inter-photon separation far exceeds the characteristic distance that a photon travels between two successive absorption events, the situation represents a memoryless, Markovian regime wherein the photons remain predominantly uncorrelated. Alternatively, when the inter-photon distance pales drastically in comparison to the aforementioned characteristic distance, the atom remains saturated upon absorbing a single photon representing a strongly non-Markovian regime. The details are furnished in the next section.

\subsection{Wave functions in the Markovian and non-Markovian limits}

In  light of the foregoing depiction of the multiphoton scattering process, let us now investigate the wave functions in the two regimes defined by the inequalities (i) $\dfrac{\Gamma}{v_g}\abs{x_i-x_j}\gg 1$ and (ii) $\dfrac{\Gamma}{v_g}\abs{x_i-x_j}\ll 1$ respectively, for any two coordinate pairs $x_i$ and $x_j$ satisfying  $x_i,x_j$$>0$.To facilitate a lucid interpretation withal, we consider the near-resonant setting ($\omega_k\approx\Omega$) and also take the large-time limit, which, in context of the relevant system timescales, can be expressed as $t\gg \dfrac{\abs{x_i}}{v_g}$, $\forall i\in\{1,2,3,\hdots,N-1\}$.  Consequently, the function $\phi(t^{(j)})$ appearing in the $j^{\text{th}}$ order term can be approximated as $\phi(t^{(j)})\approx\phi(t)$.

\begin{figure}
	\centering
	
	\begin{subfigure}{0.45\textwidth}
		\includegraphics[width=1.1\textwidth]{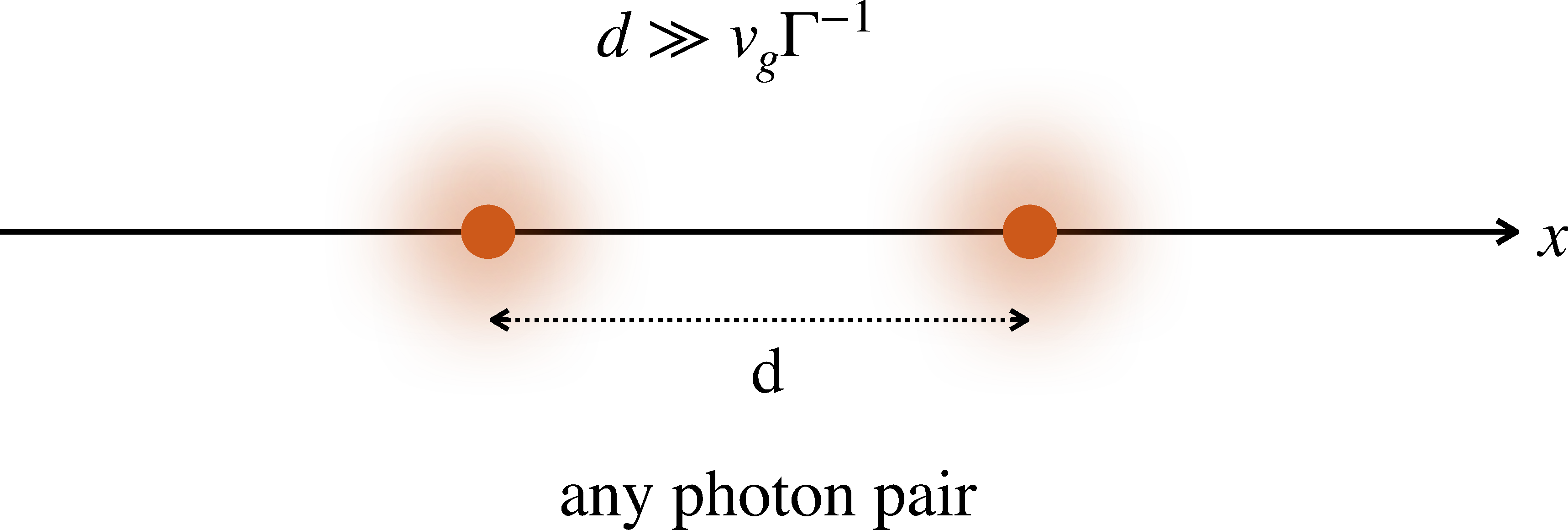}
		\caption{Markovian Limit}
		\label{figregimes1}
	\end{subfigure}
	\hfill
	\begin{subfigure}{0.49\textwidth}
		\includegraphics[width=1.1\textwidth]{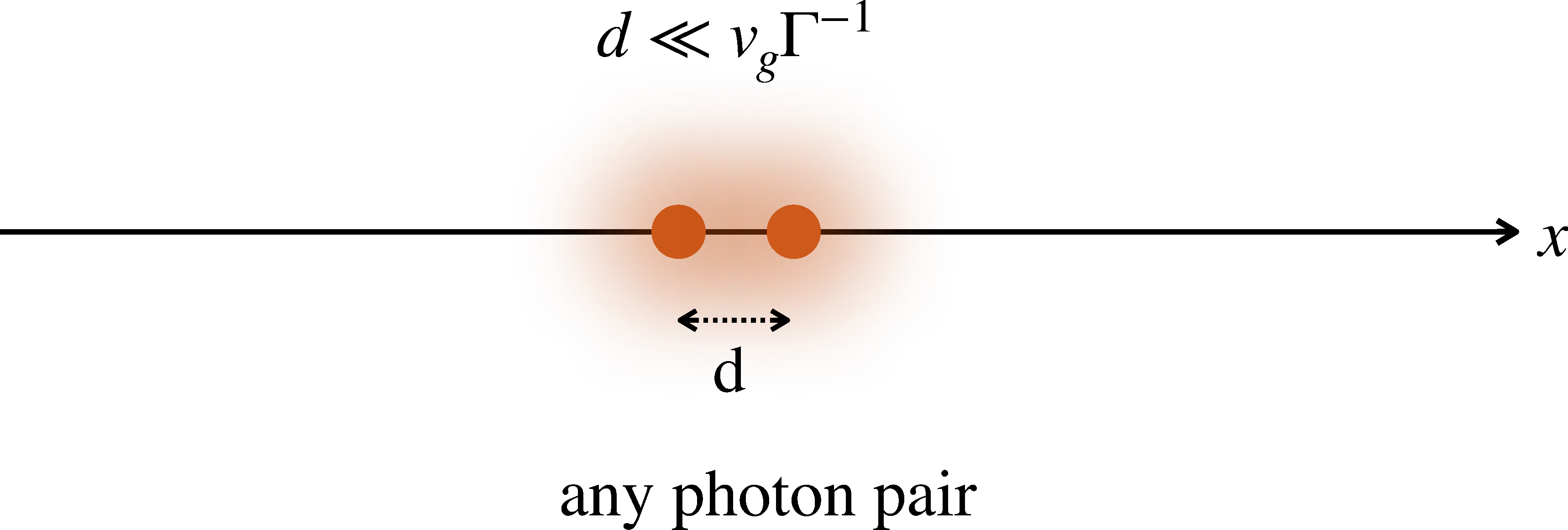}
		\caption{Non-Markovian Limit}
		\label{figregimes2}
	\end{subfigure}
	
	\caption{When the inter-photon separation $d$ is much larger than the characteristic distance that an emitted photon traverses during the atomic relaxation lifetime $\Gamma^{-1}$, it signifies a memoryless, Markov-type regime wherein the individual single-photon scattering events share no correlation with one another. Any inter-photon correlations remain strongly quenched. Conversely, when the separation becomes insignificant, the absorption of a random photon renders the atom inert to any subsequent photon absorptions (see also Table \ref{table4}). Since the atom retains full memory of its first interaction, this signifies an extreme Non-Markovian limit. Here, $d=\max_{i,j} \abs{x_i-x_j}$ represents the largest separation between any two photons.}
	\label{figregimes}
\end{figure}

\paragraph{Markovian regime:} Under the condition $\dfrac{\Gamma}{v_g}\abs{x_i-x_j}\gg 1$ (Fig. \ref{figregimes1}), we find that $\phi\bigg(\dfrac{x_{\lambda_m}^{(m)}}{v_g}\bigg)\approx -1$. Then we have
\begin{align}
 e^{(l)}(\va{x}_{N-1};t)&\approx\mathcal{A}_0\prod_{q=1}^{N-1}e^{i(kx_q-\omega_kt)}\phi(t)\Bigg[1+\Comb{l}{l-1}(\mathbf{t}_k-1)+\Comb{l}{l-2}(\mathbf{t}_k-1)^2+\hdots\Comb{l}{0}(\mathbf{t}_k-1)^l\Bigg]\notag\\
&=\mathcal{A}_0\prod_{q=1}^{N-1}e^{i(kx_q-\omega_kt)}\phi(t)[1+(\mathbf{t}_k-1)]^l\notag\\
&=[\mathbf{t}_k]^l\hspace{1mm}e^{(0)}(\va{x}_{N-1};t). \label{e24}
\end{align}
Physically, this means that the photons act as uncorrelated projectiles in this regime, with each of the $l$ photons having traversed independently across the atom without any photon absorption/emission event affecting the succeeding ones. It is, therefore, a \textit{memoryless regime} of photon scattering wherein the atom has adequate recovery time ($\sim{\abs{x_i-x_j}}/{v_g}\gg \Gamma^{-1}$) in between two absorption/emission events. We next explore the exactly opposite scenario wherein a single absorption event has the potential to block off any future absorptions.

\begin{table}[h!]
	\centering
	\begin{tabular}{ | m{19em}| m{19em} | }
		\hline
		\centering \vspace{1mm}$d\gg v_g\Gamma^{-1}$\vspace{1mm} & \vspace{1mm} \hspace{22mm} $d\ll v_g\Gamma^{-1}$\vspace{1mm}\\ [0.5ex]
		\hline\hline
		\small \vspace{2.5mm} Markov-type regime with a very short emission lifetime.\vspace{2.5mm} & \small \vspace{2.5mm} Non-Markovian regime with an exceptionally large emission lifetime. \vspace{2.5mm}\\
		\hline
		\small  \vspace{2.5mm}Supports multiple, sequential absorption events.\vspace{2.5mm} & \small \vspace{2.5mm} Admits only a single absorption event.\vspace{2.5mm}\\
		\hline
		\small \vspace{2.5mm} Absorption events are independent and pre-emitted photons remain uncorrelated.\vspace{2.5mm} & \small \vspace{2.5mm} Atom remains saturated and cannot be perturbed by subsequent photon interactions. \vspace{2.5mm}\\
		\hline
	\end{tabular}
	\vspace{1mm}
	\caption{A summary of the Markov-type and the Non-Markovian regimes with regard to multiphoton scattering in waveguide QED. As defined in Fig. \ref{figregimes}, $d=\max_{i,j} \abs{x_i-x_j}$ denotes the maximum value of the photon-photon separation.\\}
	\label{table4}
\end{table}

\paragraph{Extreme Non-Markovian limit:} In the limit $\dfrac{\Gamma}{v_g}\abs{x_i-x_j}\ll 1$ (Fig. \ref{figregimes2}), the difference in the arrival times of any two photons is much shorter than the relaxation lifetime, and we can treat terms that go as $\phi\bigg(\dfrac{x_{\lambda_m}^{(m)}}{v_g}\bigg)\ll 1$ perturbatively. This ensures that only the zeroth- and the first-order terms make dominant contributions, implying that
\begin{align}
 e^{(l)}(\va{x}_{N-1};t)&\approx\mathcal{A}_0\prod_{q=1}^{N-1}e^{i(kx_q-\omega_kt)}\phi(t)\Bigg[1+\Comb{l}{l-1}(\mathbf{t}_k-1)+\sum_{i,j}\mathcal{O}\bigg(\frac{\Gamma}{v_g}\abs{x_i-x_j}\bigg)\Theta(x_i)\Theta(x_j)\Bigg]\notag\\
&\approx\mathcal{A}_0\prod_{q=1}^{N-1}e^{i(kx_q-\omega_kt)}\phi(t)[1+(\mathbf{t}_k-1)]\notag\\
&=\mathbf{t}_k\hspace{1mm}e^{(0)}(\va{x}_{N-1};t). \label{e25}
\end{align}
This relationship between the regional scattering amplitudes testifies to the fact that intense saturation effects in the atom due to the first absorption event temporarily strips the atom of its ability to absorb any other photons. In other words, no matter how many photons run into the atom in this short timeframe ($\sim{\abs{x_i-x_j}}/{v_g}\ll \Gamma^{-1}$), the atom fails to secure the requisite recovery time to be able to re-absorb a second photon. This is a situation wherein all but a single photon gets to interface with a saturated atom. This phenomenon is essentially antithetical to the preceding Markovian scenario wherein the atom participated liberally in a cascade of independent photon absorption/emission events. For a quick reference, the identifying characteristics of these two regimes are summarized in Table \ref{table4}.

Having discussed the physical picture underlying the Fock-state scattering mechanism, we now proceed to solve the atomic excitation dynamics for more realistic quantum mechanical models of the radiation field - that of photonic pulses with a relatively narrow spectral range. Not only does the physical picture underlying the scattering mechanism remain applicable to these pulsed light fields, the algebraic machinery developed in Appendices C and D can be extended \textit{mutatis mutandis} to these wave packets. This further strengthens the potential of our toolbox, which, quite generally, could be applied to a myriad of electromagnetic pulses.

\section{Scattering dynamics: Multiphoton wave packets} \label{S4}

It is well known that plane waves, like the one considered in the preceding sections, suffer from normalizability issues since the concerned probabilities are divergent. In reality, physical electromagnetic fields are intrinsically broadened around a central frequency, and if the linewidth in frequency space is sufficiently narrow, such a field can be used to model a \textit{quasimonochromatic} plane wave. In the limit when this frequency broadening is smaller than all other physically relevant frequency scales of the system, we could closely replicate plane-wave characteristics. But more generally speaking, we can always modulate the incident plane waves by means of a smearing function to impart a finite linewidth, thereby rendering the full wave function normalizable. For instance, any $N$-photon plane-wave-type wave function as defined by Eq. \ref{e12} and weighted by a \textit{normalizable} envelope function, \textit{viz.}
\begin{align}
f(\va{x}_N)&=\xi_{\kappa}(\va{x}_N)e^{ik\sum_{j=1}^Nx_j}, \label{e26}
\end{align}
that satisfies
\begin{align}
\int_{-\infty}^{\infty}\abs{\xi_{\kappa}(\va{x}_N)}^2\dd{\va{x}_N}=1, \label{e27}
\end{align}
would represent a perfectly physical electromagnetic field. The parameter $\kappa$ could denote the linewidth parameter and then $\xi_{\kappa}(\va{x}_N)$ is termed as the smearing function. Viewed in the frequency space, the corresponding wavefunction could be written as
\begin{align}
\tilde{f}(\va{k}_N)&=\frac{1}{(2\pi)^{N/2}}\int \dd{\va{k}_N}\xi_{\kappa}(\va{x}_N)e^{-i(k_j-k)\sum_{j=1}^Nx_j}\notag\\
&=\tilde{\xi}_{\kappa}\bigg(\{k_j-k\}\bigg), \label{e28}
\end{align}
where $\tilde{f}(\va{k}_N)$ and $\tilde{\xi}_{\kappa}(\va{k}_N)$ represent the Fourier transforms of $f(\va{x}_N)$ and $\xi_{\kappa}(\va{x}_N)$ respectively. Given the practical significance of these smeared out photon pulses, it is therefore worthwhile to extend the derivation of our wave function dynamics to certain specific types of wave packets. In particular, we focus on the Lorentzian-type (in frequency space) and the Gaussian-type wave packets, which are tremendously important in optics (c. f. Fig. \ref{figplane}). We show that exact analytical solutions to the time-dependent wave functions can be worked out for both these wave packets. While the derivation for the Lorentzian pulse is contained in this section, the more complicated Gaussian case is relegated to Appendix E. However, we would mostly be concerned with narrow-band pulses that can effectively mimic single-mode Fock states of the radiation field. 

A normalized, Lorentzian-type, uncorrelated, multiphoton wave packet centered around the wave number $k$ and endowed with a linewidth $\kappa$ would be mapped into a damped plane-wave profile in position space, with the damping parameter marking any aberrations from the plane-wave behavior. The initial conditions on the full quantum system would accordingly morph into
\begin{align}
f(\va{x}_N;0)&=\bigg({\kappa}^{N/2}e^{-\kappa\sum_{j=1}^N\abs{x_j}}\bigg)e^{ik\sum_{j=1}^Nx_j},\notag\\
e(\va{x}_{N-1};0)&=0. \label{e29}
\end{align}
The initial state is perfectly normalized, i.e.,
\begin{align}
	\bra{\psi_N(0)}\ket{\psi_N(0)}=\norm{\int\dd{\va{x}_{N}}f(\va{x}_{N};0)\ket{\va{x}_{N}}}^2=\int\dd{\va{x}_{N}}\abs{f(\va{x}_{N};0)}^2=\kappa^N\bigg(\int\dd{x}e^{-2\kappa\abs{x}}\bigg)^N=1. \label{e30}
\end{align}
For sufficiently small values of the parameter $\kappa$, this represents a quasi-plane wave with its amplitude dying off as $\abs{x}\rightarrow\mathcal{O}(1/\kappa)$. In view of the modified initial conditions, the function $e_0(\va{x}_{N-1};t)$, previously defined in Eqs. \ref{e17} can be updated as
\begin{align}
e_0(\va{x}_{N-1};t)&=\sqrt{N}\mathcal{V}e^{i\Omega t}{\kappa}^{N/2}e^{-i(\omega_k-iv_g\kappa)t}\bigg(e^{i\sum_{j=1}^{N-1}(kx_j-\omega_kt)}\bigg)\bigg(e^{-\kappa\sum_{j=1}^{N-1}\abs{x_j-v_gt}}\bigg)\int_{0}^{t}\dd\tau \hspace{1mm} e^{i(\omega_k-\tilde{\Omega}-iv_g\kappa)\tau}\notag\\
&=\mathcal{A}_{\kappa}\bigg(e^{ik\sum_{j=1}^{N-1}x_j}\bigg)\bigg(e^{-\kappa\sum_{j=1}^{N-1}\abs{x_j-v_gt}}\bigg)e^{-i(N\omega_k-\Omega-iv_g\kappa)t}\bigg(e^{i(\omega_k-\tilde{\Omega}-iv_g\kappa)t}-1\bigg), \label{e31}
\end{align}
where a new coefficient $\mathcal{A}_{\kappa}=-\dfrac{\sqrt{N}i\mathcal{V}\hspace{1mm}{\kappa}^{N/2}}{\omega_k-{\Omega}+i(\Gamma-v_g\kappa)}$ has been introduced. One can proceed with the same treatment as was applied to the case of plane waves (Eq. \ref{a77} in Appendix D), and update the solution accordingly. In addition to the three lemmas laid out in Appendix C, two other lemmas are required in this context to calculate the final wave function. These are stated below.

\textit{\textbf{Lemma 4:} The vector $\va{X}^{(j)}$, with $j\in\{1,N-1\}$ has exactly $j$ coordinates which are forced to be negative by the constraints of integration in Eq. \ref{a77}, \textit{viz.}
	\begin{align*}
		x_{l}^{(j)}\bigg|_{l\in\mathbf{\Lambda}_{j}}<0.
	\end{align*}
	The rest of the coordinates can be either positive or negative.} 
	
To verify the above inequalities, note that
\begin{align}
	x_{l}^{(j)}=(x_l-x_{\lambda_{j-1}})+\sum_{n=0}^{j-1}(v_g\tau_n-x_{\lambda_n}^{(l)})\delta_{l,\lambda_n}. \label{e32}
\end{align}
Now, the integration term in which $e_0(\va{X}^{( j)}; t^{(j)})$ appears (c.f. Eq. \ref{a77}), mandates the constraints $x_{\lambda_m}^{(m)}> 0$ $\forall m\in\{0,1,2,\hdots,j-1\}$, or that $0 < x_{\lambda_0} < x_{\lambda_1} < x_{\lambda_2} < \hdots < x_{\lambda_{j-1}}$. Hence, $x_l - x_{\lambda_{j-1}} < 0$ whenever $l\in \mathbf{\Lambda}_j$, which proves Lemma 4. 

\textit{\textbf{Lemma 5:} Since $e_0(\va{X}^{( j)}; t^{(j)})$ contributes if and only if $t^{(j)}> 0$, the above lemmas yield the following:
	\begin{enumerate}
		\item $\abs{x_l^{(j)}-v_gt^{(j)}}_{l\in \mathbf{\Lambda}_j}=v_gt^{(j)}-x_l^{(j)}$.
		\item $\abs{x_l^{(j)}-v_gt^{(j)}}_{l\notin \mathbf{\Lambda}_j}=\abs{x_l-v_gt}$.
\end{enumerate}}

Using these lemmas, one can simplify the expression for the functions $e_0(\va{X}^{(j)};t^{(j)})$:
\begin{align}
	e_0(\va{X}^{(j)};t^{(j)})=\mathcal{A}\bigg(e^{ik\sum_{l=1}^{N-1}x_l}\bigg)\bigg(e^{-\kappa\sum_{l=1}^{N-1}\abs{x_l-v_gt}}\bigg)\bigg(e^{-i(Nk-\kappa)x_{\lambda_{j-1}}}\bigg)\bigg(e^{i(\omega_k-iv_g\kappa)\sum_{l=0}^{j-1}\tau_l}\bigg)\bigg(e^{-i(N\omega_k-\Omega-iv_g\kappa)t^{(j)}}\bigg)\bigg(e^{i(\omega_k-\tilde{\Omega}-iv_g\kappa)t^{(j)}}-1\bigg). \label{e33}
\end{align} 

Upon substituting these expressions and performing the integrations in Eq. \ref{a77}, we obtain the time-dependent wave function for a Lorentzian input:
  \begin{align}
e(\va{x}_{N-1};t)=\mathcal{A}_{\kappa}e^{-i(\omega_k-\Omega-iv_g\kappa)t}\bigg(\prod_{l=1}^{N-1}e^{i(kx_l-\omega_kt)}\bigg)\bigg(e^{-\kappa\sum_{j=1}^{N-1}\abs{x_j-v_gt}}\bigg)&\Bigg[\phi_{\kappa}(t)+\sum_{j=1}^{N-1}(1-\mathbf{t}_k)^j\cross\notag\\&\sum_{\mathbf{\Lambda}_j\in\mathcal{P}_j}\phi_{\kappa}(t^{(j)})\Theta(t^{(j)})\prod_{m=0}^{j-1}\phi_{\kappa}\bigg(\frac{x_{\lambda_m}^{(m)}}{v_g}\bigg)\Theta(x_{\lambda_m}^{(m)})\hspace{1mm}\Bigg]\hspace{1mm}. \label{e34}
  \end{align}

The old function $\phi(t)$ and the single-photon transmission coefficient $\mathbf{t}_k$ as applicable to the plane-wave setting have now been transformed into
\begin{align*}
\phi_{\kappa}(t)&=e^{i(\omega_k-\Omega)t}e^{-(\Gamma-v_g\kappa) t}-1,\notag\\
\mathbf{t}_k&=\dfrac{\omega_k-\Omega-i(\Gamma-v_g\kappa)}{\omega_k-\Omega+i(\Gamma-v_g\kappa)}.
\end{align*}

\section{Quantum Rabi oscillations in a waveguide configuration} \label{S5}

In this section, we call upon the dynamical solution to the Lorentzian input that we just derived and calculate the atomic excitation probability when the atom starts out in its ground state. Treating the Gaussian case is much more cumbersome and the associated algebra is highly nontrivial. On the  other hand, the result for the Lorentzian case can be processed with relative facility yielding the pertinent transition probabilities and leading, in the limit of small $\kappa$, to a nice and intuitive visualization of plane-wave scattering. As an interesting consequence of this undertaking, we also get to derive the quantum analog of Rabi oscillations for a quasimonochromatic plane wave in the strong-excitation regime. Viewed in the quantum mechanical paradigm, such a regime would correspond to the asymptotic limit of large photon numbers. In the weak excitation regime, to the contrary, a plane wave lacks the potential to induce atomic transitions to the excited state. The scenario for pulsed oscillations, however, turns out to be conspicuously different as finite-width wave packets can sharply raise the excitation amplitude, as we shall later exemplify for the cases of single-photon and two-photon pulsed fields. For the purposes of the current section, we restrict ourselves to a quasimonochromatic line source and slowly establish correspondence with the semiclassical model for Rabi oscillations. 

The analytical expression for the time-dependent wave function $e(\va{x}_{N-1};t)$ (Eq. \ref{e34}) allows us to compute the probability that the atom goes over to its excited state at any later time $t>0$. This is given by
\begin{align}
	p_N^{(e)}(t)&=\norm{\int\dd{\va{x}_{N-1}}e(\va{x}_{N-1};t)e^{-i\Omega t}\ket{\va{x}_{N-1}}}^2\notag\\
	&=\int\dd{\va{x}_{N-1}}\abs{e(\va{x}_{N-1};t)}^2\notag\\
	&=\abs{\mathcal{A}_{\kappa}}^2\int\dd{\va{x}_{N-1}}e^{-2\kappa\sum_{j=1}^{N-1}\abs{x_j-v_gt}}\Bigg[\abs{T_1}^2+T_1^*T_2+T_1T_2^*+\abs{T_2}^2\Bigg], \label{e35}
\end{align}
wherein 
\begin{align}
	\abs{T_1}^2&=\abs{\phi_{\kappa}(t)}^2,\notag\\
	T_1^*T_2&=\phi_{\kappa}^*(t)\sum_{j=1}^{N-1}(1-\mathbf{t}_k)^j\sum_{\mathbf{\Lambda}_j\in\mathcal{P}_j}\phi_{\kappa}(t^{(j)})\Theta(t^{(j)})\prod_{m=0}^{j-1}\phi_{\kappa}\bigg(\frac{x_{\lambda_m}^{(m)}}{v_g}\bigg)\Theta(x_{\lambda_m}^{(m)}),\notag\\
	\abs{T_2}^2&=\sum_{i,j=1}^{N-1}(1-\mathbf{t}_k^*)^i(1-\mathbf{t}_k)^j\sum_{\mathbf{\Lambda}_i\in\mathcal{P}_i,\mathbf{\Lambda'}_j\in\mathcal{P}_j}\phi_{\kappa}^*\bigg(t-\frac{x_{\lambda_{i-1}}}{v_g}\bigg)\phi_{\kappa}\bigg(t-\frac{x_{\lambda'_{j-1}}}{v_g}\bigg)\notag\\&\cross\Theta\bigg(t-\frac{x_{\lambda_m}^{(m)}}{v_g}\bigg)\Theta\bigg(t-\frac{x_{\lambda'_n}^{(n)}}{v_g}\bigg)\hspace{1mm}\prod_{m=0}^{i-1}\prod_{n=0}^{j-1}\phi_{\kappa}^*\bigg(\frac{x_{\lambda_m}^{(m)}}{v_g}\bigg)\phi_{\kappa}\bigg(\frac{x_{\lambda'_n}^{(n)}}{v_g}\bigg)\Theta(x_{\lambda_m}^{(m)})\Theta(x_{\lambda'_n}^{(n)})\Bigg]. \label{e36}
\end{align}
 Considering that there are four terms, we decompose the net probability as
 \begin{align}
 	p_N^{(e)}(t)=\sum_{j=0}^3p_j, \label{e37}
 \end{align} 
wherein 
\begin{align}
	p_0&=\abs{\mathcal{A}_{\kappa}}^2\int\dd{\va{x}_{N-1}}e^{-2\kappa\sum_{j=1}^{N-1}\abs{x_j-v_gt}}\abs{T_1}^2,\label{e38}\\
	p_1&=\abs{\mathcal{A}_{\kappa}}^2\int\dd{\va{x}_{N-1}}e^{-2\kappa\sum_{j=1}^{N-1}\abs{x_j-v_gt}}T_1^*T_2,\label{e39}\\
	p_2&=\abs{\mathcal{A}_{\kappa}}^2\int\dd{\va{x}_{N-1}}e^{-2\kappa\sum_{j=1}^{N-1}\abs{x_j-v_gt}}T_1T_2^*=p_1^*,\label{e40}\\
	p_3&=\abs{\mathcal{A}_{\kappa}}^2\int\dd{\va{x}_{N-1}}e^{-2\kappa\sum_{j=1}^{N-1}\abs{x_j-v_gt}}\abs{T_2}^2,\label{e41}
\end{align}
denote the various contributions. The first integral is quite easy to calculate while the integrals pertinent to the remaining terms are much more cumbersome, and thence, we do not contend with their exact expressions. However, the overall probability can be simplified in some special regimes as we discuss next. The correct behavior apropos of Rabi oscillations in an atom is recovered in these regimes of interest. The algebraic computation of these integrals in these particular cases can be found in Appendix F. In what follows, we merely state the final analytical results of our calculation. Interestingly, in two of these regimes, \textit{viz.} the weak-field approximation and the limit of large detunings, only the first integral (Eq. \ref{e38}) makes an appreciable contribution while the other integrals sparingly alter the dynamics. The scattering probability remains quite low under these conditions. It is only when the atom is driven close to resonance by a strong electromagnetic field that all of these integrals produce sizable contributions and must, therefore, be treated on an equal footing. Considering the complexity of calculating the final three integrals (Eqs. \ref{e39}-\ref{e41}), we first highlight the dynamics in the weak-scattering scenarios, followed by an elaborate discussion on the strong-pumping limit at the very end. 

\begin{figure}
        \centering
        \begin{subfigure}{.7\textwidth}
            \centering
		  \includegraphics[scale=0.2]{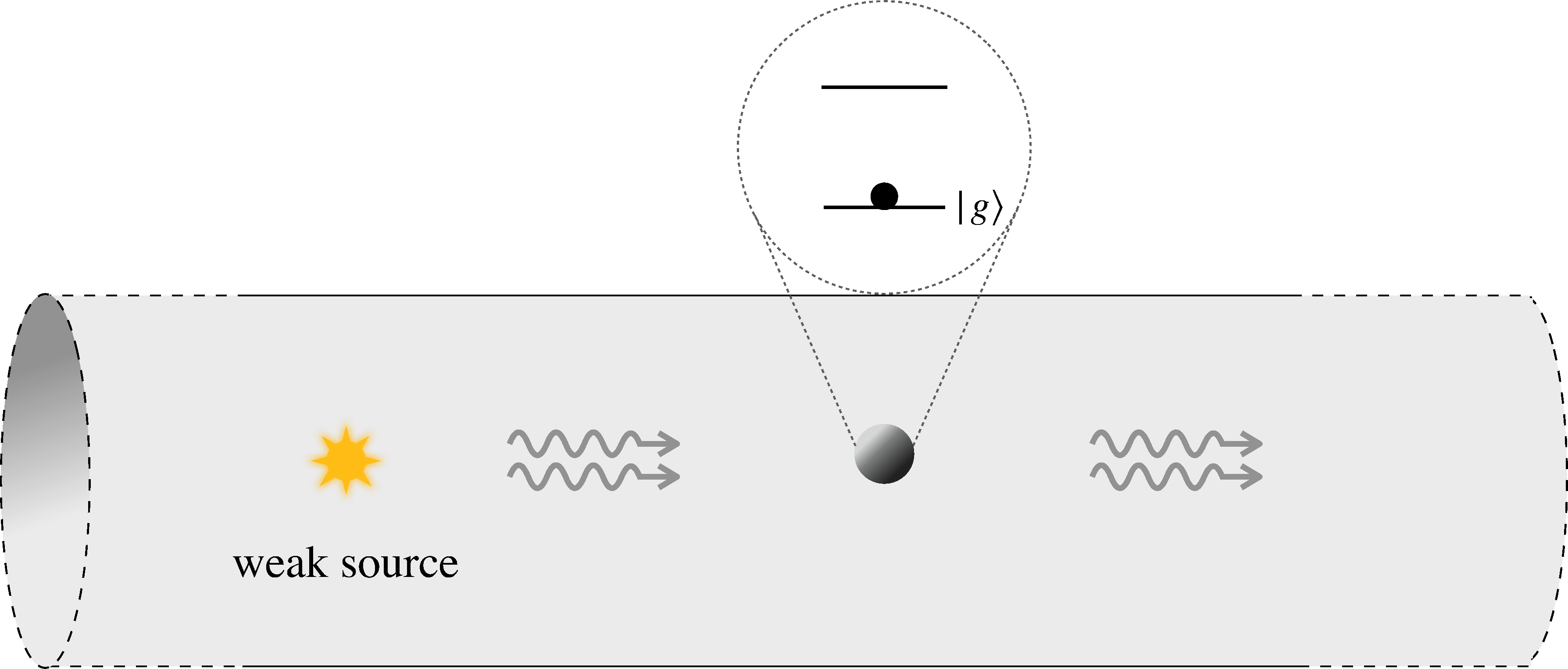}
		  \caption{Weak-pumping limit (few-photon regime) near resonance ($\omega_k\approx\Omega$).}
		  \label{figweak1}
        \end{subfigure}
        \hfill
        \begin{subfigure}{.7\textwidth}
            \centering
		  \includegraphics[scale=0.2]{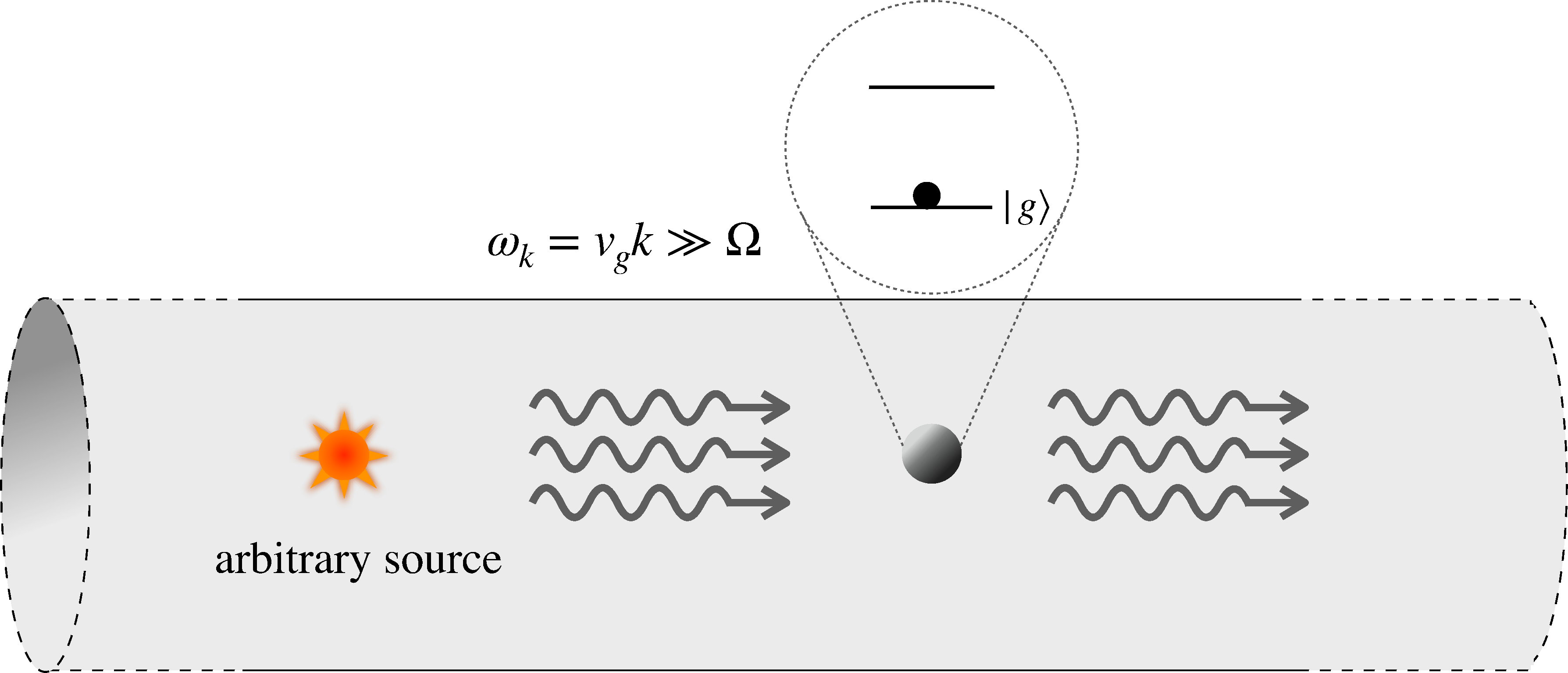}
		  \caption{Limit in which the source is far-detuned from the transition line.}
		  \label{figweak2}
        \end{subfigure} 
	 \caption{Two different regimes in which the atom-photon interaction merely acts as a perturbation and the absorption probability remains quite low. The atom remains mostly in the ground state.}
	 \label{figweak}
\end{figure}

\subsection{Near-resonant weak-pumping limit}
If $N(v_g\kappa)\ll\Gamma$, indicating low photon numbers (c.f. Fig. \ref{figweak1}), $p_0$ is the predominant contributor to the net probability. Additionally, if $t\ll\Gamma^{-1}$, the rest of the terms can be reasonably viewed as minute corrections to $p_0$. In that case, $p_N^{(e)}(t)$ reduces to
\begin{align}
p_N^{(e)}(t)\approx 2N(v_g\kappa)\Gamma t^2=Np_1^{(e)}(t), \label{e42}
\end{align}
where $p_1^{(e)}(t)$ is the excitation probability for single-photon incidence. This linear relation underscores the fact that the simple single-photon picture still holds relevance in the weak-excitation regime. Identifying the coefficient $2N(v_g\kappa)\Gamma=g^2$, we can formally map this to the standard semiclassical result in the limit of a weak Rabi frequency. In the semiclassical analysis, the weak-field excitation probability goes as 
\begin{align}
\dfrac{4g^2}{\Delta^2+4g^2}\sin^2\bigg(\sqrt{{\Delta}/{2})^2+g^2}\hspace{1mm}t\bigg)\approx g^2t^2, \label{e43}
\end{align}
where $g=\dfrac{\va{p}\vdot\va{\mathcal{E}}}{\hbar}$ is the Rabi frequency on resonance and $\Delta=\omega_k-\Omega$ the frequency detuning, with $\va{p}$ and $\va{\mathcal{E}}$ denoting the atomic transition dipole moment and the electric field amplitude respectively. Unsurprisingly, the correspondence with our result only holds water whenever the observation time remains well below the spontaneous emission lifetime of the atom. The quadratic time dependence in the scattering probability of the atom is rooted in two basic assumptions of our model, first being the supposed sharpness of the atomic transition, and second the monochromaticity of the initial wave. A well-known remedy for this is to properly account for and integrate over the width of the spectral line. This would help establish consistency with the celebrated Fermi's Golden Rule and the correct Einstein B coefficient.

\subsection{Large-detuning limit}
Physically, we anticipate that in the limit of very large frequency mismatch between the incident wave and the atomic transition line, the chances of absorption would also remain low. In addition, we expect this to remain valid for a wide range of pumping rates. Mathematically, this can be verified at once by examining the expression for the atomic excitation probability. If $\Delta\gg\Gamma$ (Fig. \ref{figweak2}), once again, just the first term, i.e., $p_0$ contributes. The rest of the terms drop out as $\mathbf{t_k}\rightarrow 1$. In this limit, $p_N^{(e)}(t)$ can be simplified as
\begin{align}
p_N^{(e)}(t)\approx \frac{8N(v_g\kappa)\Gamma}{\Delta^2}\sin^2\bigg(\frac{\Delta}{2}t\bigg). \label{e44}
\end{align}
Employing the previously identified mapping protocol $2N(v_g\kappa)\Gamma\rightarrow g^2$, we observe that this too conforms exactly with the semiclassical result, i.e., 
\begin{align}
p_N^{(e)}(t)\approx\dfrac{4g^2}{\Delta^2}\sin^2\bigg(\dfrac{\Delta}{2}t\bigg), \label{e45}
\end{align}
which holds relevance in the dispersive regime of field-matter interaction. Similar to the preceding scenario, the light field almost goes through unimpeded with the atom practically remaining in its ground state. In order to significantly enhance the excitation probability, it is therefore imperative to substantially shore up the input power of the pumping field and drive the atom close to resonance. The corresponding limiting behavior, i.e., when $N\gg 1$, is elaborated next.

\subsection{Resonant strong-pumping limit: Emergence of Rabi oscillation}

If the input field state comprises of a luxuriant supply of photons, then the contributions from $p_1$, $p_2$, and $p_3$ cannot be discarded. But most importantly, it turns out that all these terms can be expediently computed in the $N\gg 1$ limit if we neglect the dynamical effect of spontaneous emission and let $v_g\kappa\ll\Gamma$ \footnote{This approximation implicitly refers to the quasi-plane wave case.}. The calculation is further simplified upon considering the resonant setting in which we have $\omega_k=\Omega$. In that case, the first term $p_0$ remains the same as in the weak-field case, i.e.,
\begin{align}
p_0\approx 2N(v_g\kappa)\Gamma t^2=g^2t^2. \label{e46}
\end{align}
The calculation of $p_1$, $p_2$, and $p_3$ is facilitated by the linearized approximation $\phi(t)\approx -\Gamma t$ in the limit $t\ll\Gamma^{-1}$, and the details are worked out in Appendix F. The sum of these contributions is obtained to be
\begin{align}
	p_N^{(e)}(t)=p_0+p_1+p_2+p_3 \approx \sin^2(gt). \label{e47}
\end{align}
The net probability, therefore, oscillates between $0$ and $1$ with a frequency of $\dfrac{g}{2\pi}$. This completes our derivation of the quantum analogue of the $N$-photon Rabi oscillation on resonance, with the associated Rabi-frequency given by $\Omega_R=g_0\sqrt{N}$ and $g_0=\sqrt{2(v_g\kappa)\Gamma}$ being the vacuum Rabi frequency. That this expression indeed represents the quantum counterpart to the semiclassical Rabi frequency is demonstrated from first principles in Appendix G. Incidentally, this result also works fairly well when the system is approximately tuned to resonance, i.e., when $\omega_k\approx\Omega$. More precisely, there is no linear-order contribution to this expression from the detuning, in complete harmony with the semiclassical result. Quadratic-order corrections in the frequency detuning are rather nontrivial to compute analytically. 

\begin{figure}
	\captionsetup{justification=raggedright,singlelinecheck=false}
	\centering
	\includegraphics[scale=0.2]{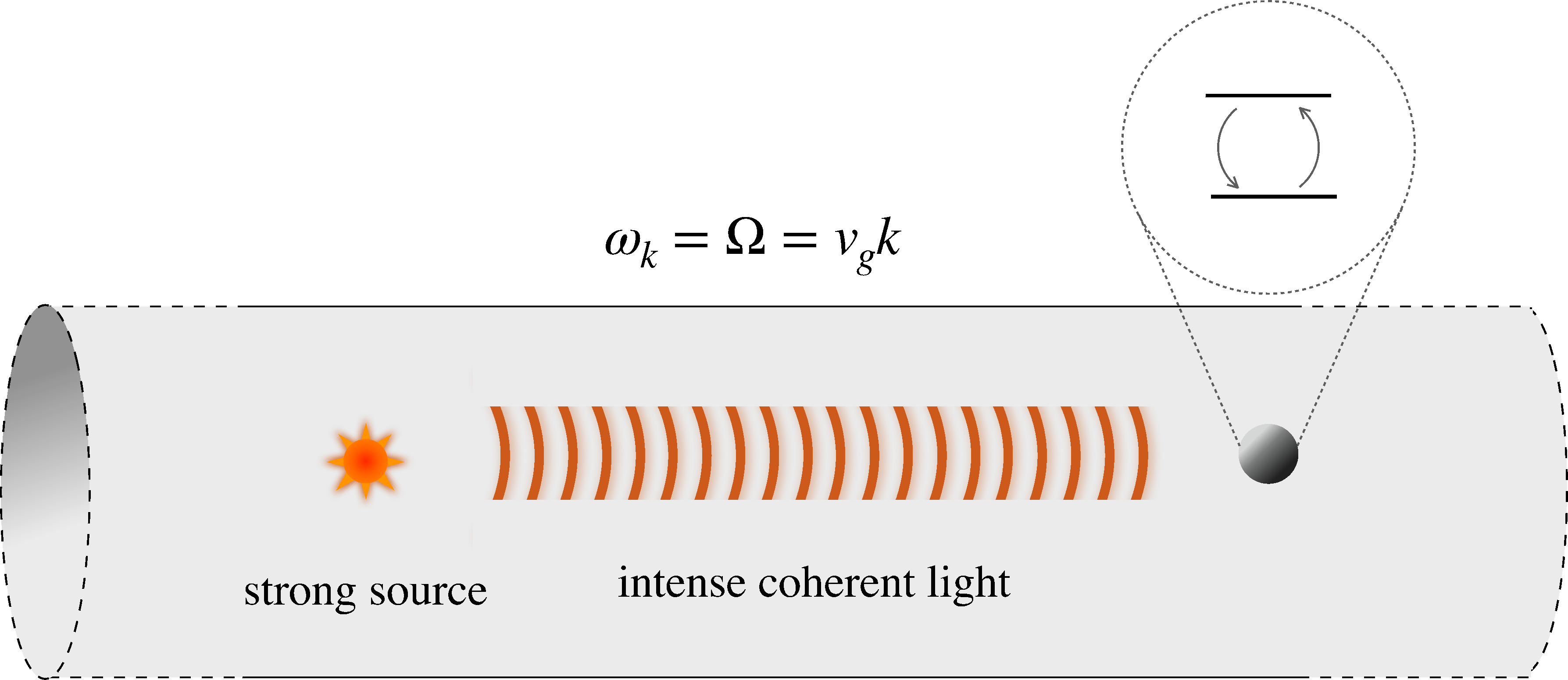}
	\caption{Resonant pumping of the atom by a strong coherent light source, which can induce oscillatory transitions between the energy levels. In the fully quantized framework, a strong pumping source could be mimicked by a multiphoton source with asymptotically large mean photon numbers. This establishes direct contact with the semiclassical predictions which also feature Rabi flopping in the atomic response.}
	\label{figrabi}
\end{figure}

Besides being in conformity with the semiclassical result, the asymptotic result of Eq. \ref{e47} also finds parity with the probabilistic behavior in a cavity setup pertaining to Fock-states of the intracavity field. That said, in the cavity-QED paradigm, the sine-squared type atomic response to Fock-state excitations holds valid even in the few-photon regime, and {ergo}, translates into collapse-and-revivial type phenomena for \textit{arbitrary} coherent-field excitations. However, the case of coherent excitation \textit{apropos of the strong pumping limit} requires more subtle considerations. For intense coherent fields of light, the atom also undergoes Rabi oscillations (c. f. Fig. \ref{figrabi}). The following discussion vindicates this claim.

For an initial coherent field state $\ket{\alpha}_k=e^{-\abs{\alpha}^2/2}\sum_{n=0}^{\infty}\dfrac{{\alpha}^n}{\sqrt{n!}}\ket{n}_k$ with a Poissonian photon number distribution $p_{\alpha}(n)=\dfrac{e^{-\abs{\alpha}^2}\abs{\alpha}^{2n}}{n!}$, the probability of atomic excitation would be obtained as
\begin{align}
p_e^{(\alpha)}(t)=\sum_{n=1}^{\infty}p_{\alpha}(n)p_e^{(n)}(t).\label{e48}
\end{align}
An exact computation of this sum is nontrivial since an analytical expression for $p_e^{(n)}(t)$ would be rather unwieldy, particularly in the few-photon regime. Nevertheless, when the mean photon number $\expval{n}_{\alpha}=\abs{\alpha}^2$  is much greater than unity, we can argue that the dominant manifolds contributing to the above sum concern photon numbers in the range $\expval{n}_{\alpha}-\sqrt{\expval{n}_{\alpha}}/2<n<\expval{n}_{\alpha}+\sqrt{\expval{n}_{\alpha}}/2$. This essentially amounts to ignoring contribution from the tail of the Poisson distribution. In this range, since $\expval{n}_{\alpha}\gg 1$, we can approximate $p_e^{(n)}(t)\approx \sin^2 (g_0\sqrt{n}t)$ and extend the range of summation to include the contribution from the tail of the photon number distribution, i.e.,
\begin{align}
p_e^{(\alpha)}(t)\approx\sum_{n=1}^{\infty}p_{\alpha}(n)\sin^2 (g_0\sqrt{n}t).\label{e49}
\end{align}
Even in the current form, summing up the aforementioned series would be virtually impossible due to the presence of the factor $\sqrt{n}$ in the argument of the sine function. Once again, the asymptotic limit comes to our rescue. Considering the fact that the distribution $p_{\alpha}(n)$ dies away quickly beyond the range $\expval{n}_{\alpha}-\sqrt{\expval{n}_{\alpha}}/2<n<\expval{n}_{\alpha}+\sqrt{\expval{n}_{\alpha}}/2$, we can expand the function $\sqrt{n}$ in a Taylor series about $\sqrt{\expval{n}_{\alpha}}$,
\begin{align}
n^{1/2}&\approx {\expval{n}_{\alpha}}^{1/2}\bigg(1+\frac{n-{\expval{n}_{\alpha}}}{{\expval{n}_{\alpha}}}\bigg)^{1/2}\notag\\
&\approx {\expval{n}_{\alpha}}^{1/2}\Bigg[1+\frac{n-{\expval{n}_{\alpha}}}{2{\expval{n}_{\alpha}}}+\mathcal{O}\bigg(\bigg(\frac{n-{\expval{n}_{\alpha}}}{{\expval{n}_{\alpha}}}\bigg)^2\bigg)\Bigg].\label{e50}
\end{align}
In the manifolds of interest spanning the full width of the photon number distribution, we have the inequality 
\begin{align}
\abs{\dfrac{n-{\expval{n}_{\alpha}}}{{\expval{n}_{\alpha}}}}<\dfrac{\sqrt{\expval{n}_{\alpha}}/2}{\expval{n}_{\alpha}}=\dfrac{1}{2\sqrt{\expval{n}_{\alpha}}},\label{e51}
\end{align}
which is vanishingly small in the asymptotic limit. Thence, upon truncating the above series at the first order, we obtain
\begin{align}
&g_0\sqrt{n}t< g_0\sqrt{{\expval{n}_{\alpha}}}t+\underbrace{\frac{1}{{2}}g_0t}_{\ll1},\notag\\
\implies &\sin(g_0\sqrt{n}t)\approx \sin( g_0\sqrt{{\expval{n}_{\alpha}}}t)+\mathcal{O}(g_0t).\label{e52}
\end{align}
Here, we have employed the inequality $g_0t=\sqrt{2(v_g\kappa)\Gamma}t\ll\Gamma t\ll 1$. Consequently, to a good approximation, the excitation probability for a strong coherent source simplifies to
\begin{align}
p_e^{(\alpha)}(t)&\approx\sum_{n=1}^{\infty}p_{\alpha}(n)\sin^2 (g_0\sqrt{{\expval{n}_{\alpha}}}t)\notag\\
&=\sin^2 (g_0\abs{\alpha}t),\label{e53}
\end{align}
in complete agreement with optical resonance experiments involving coherent laser sources. We have thus justified from purely quantum theoretic grounds that a strong coherent light field interfacing with a two-level atom elicits sinusoidal Rabi oscillations, with the Rabi frequency determined by the mean photon number in the radiation field. Notably, the mean number of photons in a quantum mechanical field also directly determines its intensity. 

\begin{figure}
	\captionsetup{justification=raggedright,singlelinecheck=false}
	\centering
	\includegraphics[scale=1]{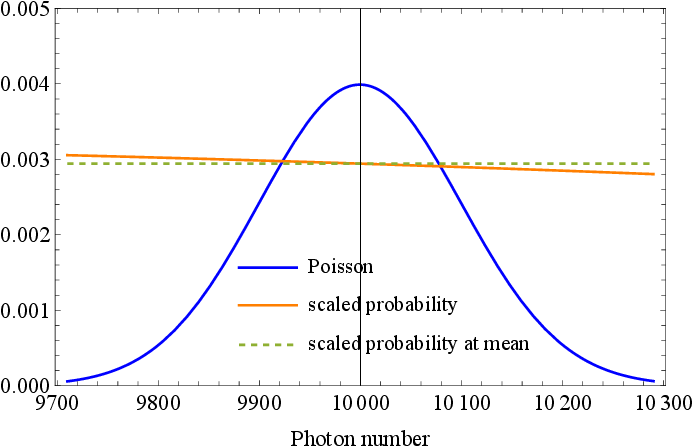}
	\caption{Graphical justification of the approximation $\sin^2(g_0\sqrt{n}t)\approx\sin^2(g_0\sqrt{\expval{n}_{\alpha}}t)$ when the mean number of photons in the coherent state is $\expval{n}_{\alpha}=10^4$. Other relevant parameter assumed is $g_0t=0.05$, where $g_0=\sqrt{2(v_g\kappa)\Gamma}$.}
	\label{figpoisson}
\end{figure} 

The relevance of our approximation is also affirmed by the plots in Fig. \ref{figpoisson}, which collectively depict the slow variation of the sinusoidal function $\sin(g_0\sqrt{n}t)$ across the width of the Poissonian distribution. It makes for a relevant observation that this approximation scheme works well only in the asymptotic limit ${\expval{n}_{\alpha}}\gg 1$, which establishes contact with the canonical operating regime of semiclassical optics. To summarize the key findings of this section, we tabulate, in Table \ref{table5}, the limiting behaviors of the atomic excitation probability in the three important regimes discussed thereof. In the next section, we consider the scattering of photon pulses with finite bandwidths.

\begin{table}[t!]
	\centering
	\begin{tabular}{ | m{9em} | m{9em}| m{9em}|  }
		\hline
		& \centering \vspace{2.5mm} Semiclassical model \vspace{2.5mm} & \vspace{1mm} \hspace{2mm} Quantum treatment \vspace{2.5mm}\\ [0.5ex]
		\hline\hline
		\small \vspace{2.5mm} \centering Near-resonant weak-field limit \vspace{2.5mm} & \small \centering \vspace{2.5mm} $g^2t^2$\vspace{2.5mm}  & \vspace{2.5mm}\hspace{9mm}\small $2{\expval{n}_{\alpha}}(v_g\kappa)\Gamma t^2$\vspace{2.5mm}   \\
		\hline
		\small \vspace{2.5mm}\centering Large-detuning limit (arbitrary field strength) \vspace{2.5mm} & \centering \vspace{2.5mm} \small $\frac{4g^2}{\Delta^2}\sin^2(\frac{\Delta}{2}t)$ \vspace{2.5mm} & \vspace{2.5mm} \small \hspace{4mm} $\frac{8{\expval{n}_{\alpha}}(v_g\kappa)\Gamma}{\Delta^2}\sin^2(\frac{\Delta}{2}t)$ \vspace{2.5mm} \\
		\hline
		\small \vspace{2.5mm}\centering Near-resonant strong-field limit \vspace{2.5mm} & \small \vspace{2.5mm}\centering $\sin^2(gt)$ \vspace{2.5mm} & \vspace{2.5mm}\hspace{2mm}\small $\sin^2\bigg(\sqrt{2{\expval{n}_{\alpha}}(v_g\kappa)\Gamma}\hspace{0.8mm}t\bigg)$ \vspace{2.5mm}\\
		\hline
	\end{tabular}
	\caption{Quantum-classical correspondence between the quantum mechanical, waveguide-QED model and the semiclassical model, which identifies the quantum mechanical Rabi frequency as $\sqrt{2{\expval{n}_{\alpha}}(v_g\kappa)\Gamma}$. Here ${\expval{n}_{\alpha}}=\abs{\alpha}^2$ denotes the mean photon number of a coherent light source.}
	\label{table5}
\end{table}

\section{Dynamics of pulsed excitations: few-photon regimes} \label{S6}
All of our probability calculations heretofore were largely simplified by the quasi-plane-wave approximation in relation to the radiation source. For pulses with finite bandwidth, i.e., when $v_g\kappa\sim\mathcal{O}(\Gamma)$, the complexity of the probability calculations gets significantly amplified. This is because as the photon number keeps growing, many of the simplifications invoked earlier break down. However, in the few-photon regime, exact answers can be derived by brute-force or numerical integrations. In this section, we briefly discuss a couple of instances where broadband few-photon pulses (assumed Lorentzian - c.f. Eq. \ref{e29})) bear the potential to induce atomic transitions with higher probabilities.

Let us consider the single-photon scenario first, in which case, the probability of excitation is given simply by 
\begin{align}
p^{(1)}&=\frac{2{(v_g\kappa)\hspace{1mm}\Gamma}\hspace{1mm}e^{-2(v_g\kappa)t}\hspace{1mm}\abs{\phi_{\kappa}(t)}^2}{(\omega_k-\Omega)^2+(\Gamma-v_g\kappa)^2}.\label{e54}
\end{align}
The corresponding scattering probability in the two-photon case is obtained as
\begin{align}
	p^{(2)}=\frac{4{(v_g\kappa)\hspace{1mm}\Gamma}\hspace{1mm}e^{-2(v_g\kappa)t}}{(\omega_k-\Omega)^2+(\Gamma-v_g\kappa)^2}\Bigg[&\abs{\phi_{\kappa}(t)}^2+\bigg\{\kappa\phi_{\kappa}^*(t)\bigg(1-\mathbf{t}_k\bigg)\int_0^{v_gt}\dd{x}\hspace{1mm}e^{-2\kappa(v_gt-x)}\phi_{\kappa}(x/v_g)\hspace{1mm}\phi_{\kappa}(t-x/v_g)+\hspace{1mm}\text{c.c.}\bigg\}\notag\\
	&+\kappa\abs{1-\mathbf{t}_k}^2\int_0^{v_gt}\dd{x}e^{-2\kappa(v_gt-x)}\abs{\phi_{\kappa}(x/v_g)}^2\abs{\phi_{\kappa}(t-x/v_g)}^2\Bigg].\label{e55}
\end{align}

\begin{figure}
	\captionsetup{justification=raggedright,singlelinecheck=false}
	\centering
	\includegraphics[scale=1.2]{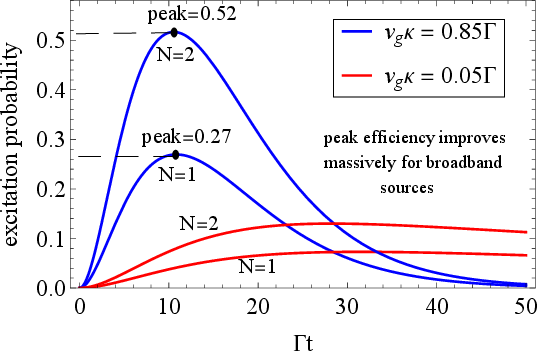}
	\caption{Time evolution of the atomic excitation probability upon illumination by few-photon ($N=1$ and $N=2$) Lorentzian wave packets with varying linewidths. The case $v_g\kappa=0.05\Gamma$ (marked in red), which describes a quasimonochromatic wave packet, is associated with a very weak excitation efficiency, whereas the broadband pulse with $v_g\kappa=0.85\Gamma$ (marked in blue) can eke out a spectacular improvement in the latter.}
	\label{pulse}
\end{figure}  

The first term ($\propto \abs{\phi_{\kappa}(t)}^2$) in $p^{(2)}$ simply equals $2p^{(1)}$ and denotes the interference-free contribution. The second term (within the curly brackets) encodes the interference between the single-photon absorption scenario and the higher-order event that a new photon could be absorbed by the atom even after one of the photons has been scattered off. The final term encodes the interference between two independent higher-order excitation pathways owing to the indistinguishability of the photons. We plot both these expressions (Eqs. \ref{e54}-\ref{e55}) in Fig. \ref{pulse}, as contextualized in a resonant setting, i.e., $\omega_k=\Omega$. Clearly, a broadband pulse can significantly improve the efficiency of atomic excitation, even down to the single-photon level. Furthermore, the two-photon scenario expectedly shows a more pronounced enhancement in the peak probability. It is, of course, worth contemplating about the origin of this improvement. Since the atomic emission line has a natural linewidth of $\Gamma$, it is not surprising that a larger bandwidth of radiation, with the width comparable in magnitude to $\Gamma$, would entail more efficient coupling with the atomic transition. As the concerned illumination source consists of photons distributed over a spectral range of $\mathcal{O}(v_g\kappa)$, this condition roughly translates to $v_g\kappa\sim\Gamma$. Under this condition, the atom has considerably higher chances of interacting with a given photon owing to its inherently broad frequency distribution, as compared to the strictly single-mode, monochromatic limit $v_g\kappa\ll\Gamma$. At the same time, it is imperative to keep in mind that the width of the illuminating pulse cannot be made indefinitely large and must be limited in size by the natural linewidth of the coupled system. For even broader pulse profiles, i.e., $v_g\kappa>\Gamma$, the theoretical applicability of the rotating wave approximation, that hinges on the condition of the light field being near-resonant with the dipolar transition, would be necessarily jeopardized.

These preliminary results are strongly suggestive of the profitability of employing broadband pulses in qubit preparations. Furthermore, while we have only probed the excitation dynamics due to Lorentzian pulses, it would be interesting to see how appropriate pulse-engineering could boost the excitation efficiency even further or serve as a useful resource in the controlled generation of qubit states. Given the ever-growing importance of waveguide-based photonic circuitry and the multitude of scalable platforms that facilitate experimental implementation, this could be a subject of future investigation. 

\section{Conclusion and Outlook} \label{S7}
In summary, we have introduced a detailed framework based on the real-space formalism in waveguide QED to analyze the scattering dynamics due to an arbitrary multiphoton wave packet incident on a two-level atom. The unitarily evolving state of the coupled system exhibits strong interference effects due to a myriad of scattering pathways available to an incident multiphoton Fock state. The zeroth order excitation pathway correctly explains the physics in the few-photon regime. As it is, it agrees perfectly with the semiclassical prediction in the limit of a weak drive and can be made consistent with Fermi's Golden Rule withal. This can be accomplished by properly incorporating the linewidth of the purportedly discrete atomic transition. The higher-order scattering events, which encode the possibility of cascaded photon absorption and emission, become more impactful as the incident photon number goes up. Incidentally, these higher-order processes also contain the fingerprints of an effective photon-photon interplay induced upon scattering. The scattered photons proceed onward as a many-body correlated bunch which shores up the complexity of the overall interference phenomenon, particularly as we stray further away from the single-photon sector. In the asymptotic limit of exceptionally large photon numbers, all excitation orders are of comparable importance and interfere strongly with each other to reproduce the anticipated hallmarks of resonant Rabi flopping. We also explore the dynamical characteristics ensuing from pulsed, few-photon wave packets and gain insights into the advantage that broadband radiation sources might afford in comparison to quasimonochromatic waves. As token specimens, we numerically simulated the results of single-photon and two-photon scattering dynamics and highlighted a prodigious enhancement in the atomic excitation probability facilitated by a frequency-broadened excitation source. 

Considering the burgeoning scientific interest in harnessing nonclassical Fock states for quantum networking protocols, as well as correlated multiphoton sources, such as photonic bound states, which arise organically in waveguide-based platforms, we believe that our detailed formulation for predicting multiphoton dynamics would find vital applications in the empirical world. Given that the scattering wavefunction in the few-photon regime bears commensurately few interference terms, it should be a relatively simple exercise to simulate and ratify the associated dynamical features. Since our analysis can cater to arbitrary incident pulses, it would be worthwhile to explore and utilize the full advantage of broadband few-photon radiation fields for useful quantum state preparations. On the theoretical frontier, it should be interesting to generalize our analysis to the case of multilevel scattering sources such as a three-level atom, or even, multimode input radiation. Essentially, our model could serve as a useful precursor for dealing with more nontrivial scattering problems with increased algebraic complexity. Another interesting aspect to investigate would be the quantum-classical crossover in the asymptotic limit (i.e., $N\rightarrow\infty$) of a pulsed pumping field. It is a well-known semiclassical result that pulsed laser beams with a narrow temporal profile beget a generalized Rabi flopping dynamics in the illuminated sample as dictated by the pulse area of the field, a statement also termed the pulse area theorem. Working out the quantum extension to this theorem within a fully quantum photonic framework would be a formidable but an extremely rewarding exercise.

\section{Acknowledgments}
The authors acknowledge financial support from the Chan Zuckerberg Initiative (2020-225832) and the National Science Foundation (1838996).

\section{Appendix A: Algebraic properties of the states $\bigg\{\ket{\va{x}_n}\bigg\}$} \label{sa}

The entire class of continuous-variable states $\bigg\{\ket{\va{x}_n}\bigg\}$, with $x_i\in(-\infty,\infty)$ $\forall i\in\{1,2,3,\hdots,n\}$, form a complete, mutually orthogonal set in the $n$-photon Hilbert space, up to permutation symmetry in the coordinates. The associated characteristics of this class of states are listed below with proofs.

\textit{\textbf{Statement 1:} $\ket{\va{x}_n}$ is an eigenstate of the multimode photon-number operator $\mathcal{N}_f=\int\dd{x}c^{\dagger}(x)c(x)$ with eigenvalue $n$.}

\textit{Proof:} Recall the definition $\ket{\va{x}_n}=\dfrac{1}{\sqrt{n!}}\bigg(\prod_{j=1}^nc^{\dagger}(x_j)\bigg)\ket{0}$. They key to simplifying the product $\mathcal{N}_f\ket{\va{x}_n}$ lies in resolving this into normally-ordered expressions via the repeated application of the commutation relations $[c(x),c^{\dagger}(x')]=\delta(x-x')$ and $[c^{\dagger}(x),c^{\dagger}(x')]=0$. Noting that
\begin{align}
	c(x)\prod_{j=1}^nc^{\dagger}(x_j)=\sum_{i=1}^n\delta(x-x_i)\prod_{j\neq i}^nc^{\dagger}(x_j)+\bigg(\prod_{j=1}^nc^{\dagger}(x_j)\bigg)c(x),\label{a60}
\end{align}
we find that
\begin{align}
	\mathcal{N}_f\ket{\va{x}_n}&=
	\sum_{i=1}^n\Bigg[\underbrace{\dfrac{1}{\sqrt{n!}}\int\dd{x}c^{\dagger}(x)\delta(x-x_i)\prod_{j\neq i}^nc^{\dagger}(x_j)\ket{0}\Bigg]}_{=\ket{\va{x}_n}}+\dfrac{1}{\sqrt{n!}}\prod_{j=1}^nc^{\dagger}(x_j)\int\dd{x}\underbrace{c(x)\ket{0}}_{=0}\notag\\
	&=n\ket{\va{x}_n}.
\end{align}

\textit{\textbf{Statement 2:} The states $\bigg\{\ket{\va{x}_n}\bigg\}$ are mutually orthogonal up to permutation symmetry.}

\textit{Proof:} From Eq. \ref{a60}, it is easy to see that
\begin{align}
	c(x'_1)\ket{\va{x}_n}=\frac{1}{\sqrt{n}}\sum_{i=1}^n\delta(x'-x_i)\ket{x_1,x_2,\hdots,x_{i-1},x_{i+1},\hdots,x_n},
\end{align}
and subsequently
\begin{align}
	c(x'_1)c(x'_2)\ket{\va{x}_n}=\frac{1}{\sqrt{n(n-1)}}\sum_{i_1=1}^n\sum_{i_2\neq i_1}\delta(x'_1-x_{i_1})\delta(x'_2-x_{i_2})\ket{x_1,x_2,\hdots,x_{i_1-1},x_{i_1+1},\hdots,x_{i_2-1},x_{i_2+1},\hdots,x_n}.
\end{align}
Thus, generally, the following relation would hold:
\begin{align}
	\bigg(\prod_{i=1}^nc(x_i')\bigg)\ket{\va{x}_n}&=\bigg(\frac{1}{\sqrt{n!}}\sum_{\mathcal{P}}\delta(\va{x}_n'-\mathcal{P}\va{x}_n)\bigg)\ket{0}\notag\\
	\implies \bra{\va{x}_n}\ket{\va{x}_n'}&=\dfrac{1}{n!}\sum_{\mathcal{P}}\delta(\va{x}_n'-\mathcal{P}\va{x}_n),
\end{align}
where $\mathcal{P}\va{x}_n$ represents any permutation of the coordinates of $\va{x}_n=(x_1,x_2,x_3,\hdots,x_{n})$.

\section{Appendix B: Formal solutions to the coupled-wave equations}\label{sb}
In order to formally solve Eq. \ref{e8}, we substitute the Fourier-space expansions
\begin{align}
	f(\va{x}_N;t)&=\frac{1}{(2\pi)^{N/2}}\int\dd\va{k}_N\hspace{1mm}\tilde{f}(\va{k}_N;t)e^{i\sum_{j=1}^{N}k_jx_j},\notag\\
	e(\va{x}_{N-1};t)&=\frac{1}{(2\pi)^{(N-1)/2}}\int\dd\va{k}_{N-1}\hspace{1mm}\tilde{e}(\va{k}_{N-1};t)e^{i\sum_{j=1}^{N-1}k_jx_j},\end{align}

to obtain 
\begin{align}
	\bigg(\frac{\partial}{\partial t}+iv_g\sum_{j=1}^Nk_j\bigg)\tilde{f}(\va{k}_N;t)=-\frac{i\mathcal{V}e^{-i\Omega t}}{\sqrt{2\pi N}}\bigg[\tilde{e}(k_2,k_3,\hdots,k_N;t)+\tilde{e}(k_1,k_3,\hdots,k_N;t)+\hdots\tilde{e}(k_1,k_2,\hdots,k_{N-1};t)\bigg].
\end{align}
Upon integrating over the time variable, we are led to the solution
\begin{align}
	\tilde{f}(\va{k}_N;t)=\tilde{f}(\va{k}_N;0)-\frac{i\mathcal{V}e^{-iv_g\sum_{j=1}^Nk_j t}}{\sqrt{2\pi N}}\bigg[&\int_0^t\dd{t'}\tilde{e}(k_2,k_3,\hdots,k_N;t')e^{-i(\Omega-v_g\sum_{j=1}^Nk_j)t'}+\notag\\
	&\int_0^t\dd{t'}\tilde{e}(k_1,k_3,\hdots,k_N;t')e^{-i(\Omega-v_g\sum_{j=1}^Nk_j)t'}+\notag\\
	&\hdots\int_0^t\dd{t'}\tilde{e}(k_1,k_2,\hdots,k_{N-1};t')e^{-i(\Omega-v_g\sum_{j=1}^Nk_j)t'}\bigg].
\end{align}
The Fourier inversion of the above solution turns out to be
\begin{align}
	f(\va{x}_N;t)=f\bigg(\{x_i-v_gt\};0\}\bigg)-\frac{i\mathcal{V}}{\sqrt{N}v_g}\sum_{j=1}^N\int_0^t\dd{t'}e^{-i\Omega (t-t')}\delta(x_j-v_gt')\hspace{1mm}e(\{x_i-v_gt'\}_{i\neq j};t-t'),
\end{align}
which reduces to Eq. \ref{e13}. An identical treatment of Eq. \ref{e9} in the Fourier domain yields the corresponding formal solution for $e(\va{x}_{N-1};t)$ (c. f. Eq. \ref{e15}).

\section{Appendix C: Iterative scheme for solving the wave equations} \label{sc}

Here we lay the foundation of an abstract algebraic formulation that would ultimately enable us to compute $e(\va{x}_{N-1};t)$ in terms of the known function in Eq. \ref{e17}. To that end, let us define a set $S_0=\bigg\{(\va{X}^{(0)};t^{(0)}), \lambda_0, \tau_0\bigg\}$, where $\va{X}^{(0)}=(x_1,x_2,\hdots,x_{N-1})$ is an array of $N-1$ spatial coordinates, $t^{(0)}=t$ stands for the time parameter, $\tau_0$ is a \textit{dummy} variable with the dimension of time, and $\lambda_0$ is a discrete dimensionless variable which can take integer values ranging from $1$ to $N-1$. We now introduce a linear transformation on this set that maps $S_0$ to a new set $S_1=\bigg\{(\va{X}^{(1)};t^{(1)}), \lambda_1, \tau_1 \bigg\}$ with the new elements defined as follows:
\begin{align}
	\va{X}^{(1)}&=\va{X}^{(0)}-x_{\lambda_0}^{(0)}\va{K}+(v_g\tau_0-x_{\lambda_0}^{(0)})\va{V}^{(0)},\notag\\
	t^{(1)}&=t^{(0)}-\frac{x_{\lambda_0}^{(0)}}{v_g}.\label{a65}
\end{align}
Here $\va{V}^{(0)}=(0,0,...0,\underbrace{1}_{\lambda_0^{th} index},0,...,0)$ and $\va{K}=(1,1,\hdots,1)$ are $(N-1)$-dimensional vectors, while $\tau_1$ is a new dummy variable with the dimension of time, and $\lambda_1\neq\lambda_0$ is another dummy variable taking integer values from $1$ to $N-1$. Therefore, while the transformations from $(\va{X}^{(0)};t^{(0)})$ to $(\va{X}^{(1)};t^{(1)})$ are well-defined and unique, the transformation $\lambda_0\rightarrow\lambda_1$ is solely constrained by the requirement that $\lambda_1$ must be distinct from $\lambda_0$. However, there is no requirement imposed on the transformation $\tau_\rightarrow\tau_0$, and they can be interpreted as representing two independently selected timelike variables. The transformed variable $\va{X}^{(1)}$ as underscored by Eqs. \ref{a65} can be fully spelled out as $\va{X}^{(1)}=(x_1-x_{\lambda_0},x_2-x_{\lambda_0},...,x_{\lambda_0-1}-x_{\lambda_0},\underbrace{v_g\tau_0-x_{\lambda_0}}_{\lambda_0^{th} index},x_{\lambda_0+1}-x_{\lambda_0},...,x_{N-1}-x_{\lambda_0})$. Clearly, then, in the terminology of these new notations, we can rewrite the definition of $\mathcal{T}_j$ (c. f. Eqs. \ref{e17}) as
\begin{align}
	\mathcal{T}_{\lambda_0}(\va{X}^{(0)};t^{(0)})=\int_0^{\frac{x_{\lambda_0}^{(0)}}{v_g}}\dd\tau_0 e^{-i\tilde{\Omega}\tau_0}e(\va{X}^{(1)};t^{(1)}),
\end{align}
since all the involved variables have the correct dimensional attributes. More generally, in view of the specific form of the mapping introduced, we can now define an entire hierarchy of sets $\{S_0, S_1, S_2,...S_{N-2}\}$, such that the primordial transformation from the set $S_j=\bigg\{(\va{X}^{(j)};t^{(j)}), \lambda_j, \tau_j\bigg\}$ to $S_{j+1}=\bigg\{(\va{X}^{(j+1)};t^{(j+1)}), \lambda_{j+1}, \tau_{j+1}\bigg\}$, under the same mapping, could be encoded as
\begin{align}
	\va{X}^{(j+1)}&=\va{X}^{(j)}-x_{\lambda_j}^{(j)}\va{K}+\bigg(v_g\tau_j-x_{\lambda_j}^{(j)}\bigg)\va{V}^{(j)},\notag\\
	t^{(j+1)}&=t^{(j)}-\frac{x_{\lambda_j}^{(j)}}{v_g},\label{a67}
\end{align}
wherein $x_{\lambda_j}^{(j)}$ is the $\lambda_j^{th}$ component of $\va{X}^{j}$, and $\va{V}^{(j)}=(0,0,...0,\underbrace{1}_{\lambda_j^{th} index},0,...,0)$. As before,  $\tau_{j+1}$ is a dummy timelike variable, while $\lambda_{j+1}$ is a discrete integer variable ranging from $1$ to $N-1$. The sole constraint on the parameters $\{\lambda_0,\lambda_1,\lambda_2,...\lambda_{N-2}\}$ is that no two of these variables are permitted to be equal. More precisely, the set $\{\lambda_0,\lambda_1,\lambda_2,...,\lambda_{j-1}\}$ represents a specific permutation of any $j$ integers chosen from $\{1,2,3,...,N-1\}$. For brevity, we denote $\mathbf{\Lambda}_j=(\lambda_0,\lambda_1,\lambda_2,...,\lambda_{j-1})$, and define $\mathcal{P}_j$ as the exhaustive set of all possible selections of any $j$ integers from $\{1,2,3,...,N-1\}$. Therefore, $\mathbf{\Lambda}_j\in \mathcal{P}_j$, and there are $\Perm{N-1}{j}=\dfrac{(N-1)!}{(N-1-j)!}$ such choices available. It is also worthwhile to note that the possibility of appending yet another distinct integer from $\{1,2,3,...,N-1\}$ to the set $\mathbf{\Lambda}_j$ is exhausted at $j=N-1$, thereby precluding the existence of a meaningful transformation $\mathbf{\Lambda}_j\rightarrow\mathbf{\Lambda}_{j+1}$ once $j$ increases to $N-1$. This explains why we capped off the hierarchy of sets at $S_{N-2}.$

Now considering the specified hierarchy of the above sets and the associated mapping mechanism, we now generalize Eq. \ref{e17} to
\begin{align}
	\mathcal{T}_{\lambda_j}(\va{X}^{(j)};t^{(j)})=\int_0^{\frac{x_{\lambda_j}^{(j)}}{v_g}}\dd\tau_j e^{-i\tilde{\Omega}\tau_j}e(\va{X}^{(j+1)};t^{(j+1)}),\label{a68}
\end{align}
Such recursively defined functions will be the key to obtaining the full solution to $e(\va{x}_{N-1};t)$. But before we construct the full solution by exploiting these recursion relations, let us underscore a few useful lemmas below.\\

\textit{\textbf{Lemma 1:}
	 $x_{\lambda_{j+1}}^{(j+1)}=x_{\lambda_{j+1}}-x_{\lambda_j}$ and $\sum_{l=0}^{j+1}x_{\lambda_l}^{(l)}=x_{\lambda_{j+1}}$, for $j\in\{0,N-2\}$.}

\textit{\textbf{Lemma 2:} The general vector $\va{X}^{(j+1)}$ can be reduced in terms of the starting vector $\va{X}^{(0)}$ as
\begin{align*}
	\va{X}^{(j+1)}=\va{X}^{(0)}-x_{\lambda_j}\va{K}+\sum_{l=0}^j(v_g\tau_l-x_{\lambda_l}^{(l)})\va{V}^{(l)}, \text{for} j\in\{0,N-2\},
\end{align*}
and thereby, its corresponding elements can be written as
\begin{align*}
	{x}_i^{(j+1)}=(x_i-x_{\lambda_j})+\sum_{l=0}^j(v_g\tau_l-x_{\lambda_l}^{(l)})\delta_{i,\lambda_l}.
\end{align*}}

\textit{\textbf{Lemma 3:} ${t}^{(j+1)}=t-\dfrac{x_{\lambda_j}}{v_g}$ for $j\in\{0,N-2\}$.}\\

The above corollaries can be straightforwardly proved by the principle of mathematical induction starting from the defining relations in Eq. \ref{a67}.

\section{Appendix D: Detailed derivation of Eq. \ref{e18}} \label{sd}
We now recast all the expressions in terms of the relevant set-theoretic notations involving the indexed space-time coordinates $\{\va{X}^{(j)};t^{(j)}\}$, as defined in the preceding appendix. Recalling that the observation point $(\va{x}_{N-1},t)$ is represented by the zero-indexed set $\{\va{X}^{(0)};t^{(0)}\}$, Eq. \ref{e16} now resembles
\begin{align}
	e(\va{X}^{(0)};t^{(0)})=e_0(\va{X}^{(0)};t^{(0)})-\frac{\mathcal{V}^2}{v_g}\sum_{\lambda_0=1}^{N-1}e^{i(\Omega/v_g)x_{\lambda_0}}\mathcal{T}_{\lambda_0}(\va{X}^{(0)};t^{(0)})\Theta(x_{\lambda_0}^{(0)})\Theta(t^{(1)}),\label{a69}
\end{align}
wherein we have invoked the set-mapping protocol of Appendix C to denote the retarded time $t-\dfrac{x_j}{v_g}$ as $t^{(1)}$. Note that any term under the summation in the above expansion contributes if and only if $x_{\lambda_0}^{(0)}>0$. For instance, if the coordinate $x_{4}^{(0)}<0$, then the corresponding term with $\lambda_0=4$ would not contribute. Given the simplicity of the set transformation properties introduced in Appendix C, it is compelling to try out an iterative technique to recursively simplify the terms involving the $\mathcal{T}_{\lambda_0}(\va{X}^{(0)};t^{(0)})$ functions. This is because, the function $\mathcal{T}_{\lambda_j}(\va{X}^{(j)};t^{(j)})$ is defined entirely in terms of $e(\va{X}^{(j+1)};t^{(j+1)})$, as per the definition in Eq. \ref{a68}. We can, therefore, employ this defining relation to express Eq. \ref{a69} as
\begin{align}
	e(\va{X}^{(0)};t^{(0)})=e_0(\va{X}^{(0)};t^{(0)})-\frac{\mathcal{V}^2}{v_g}\sum_{\lambda_0=1}^{N-1}e^{i(\Omega/v_g)x_{\lambda_0}^{(0)}}\Theta(x_{\lambda_0}^{(0)})\Theta(t^{(1)})\int_0^{\frac{x_{\lambda_0}^{(0)}}{v_g}}\dd\tau_0 e^{-i\tilde{\Omega}\tau_0}e(\va{X}^{(1)};t^{(1)}).\label{a70}
\end{align}
After this, we can insert the expansion for $e(\va{X}^{(1)};t^{(1)})$ by re-invoking Eq. \ref{a69}, i.e.,
\begin{align}
	e(\va{X}^{(1)};t^{(1)})=e_0(\va{X}^{(1)};t^{(1)})-\frac{\mathcal{V}^2}{v_g}\sum_{\lambda_1=1}^{N-1}e^{i(\Omega/v_g)x_{\lambda_1}^{(1)}}\mathcal{T}_{\lambda_1}(\va{X}^{(1)};t^{(1)})\Theta(x_{\lambda_1}^{(1)})\Theta(t^{(2)}),
\end{align}
into Eq. \ref{a70}, yielding
\begin{align}
	e(\va{X}^{(0)};t^{(0)})=&e_0(\va{X}^{(0)};t^{(0)})-\frac{\mathcal{V}^2}{v_g}\sum_{\lambda_0=1}^{N-1}e^{i(\Omega/v_g)x_{\lambda_0}}\Theta(x_{\lambda_0}^{(0)})\Theta(t^{(1)})\int_0^{\frac{x_{\lambda_0}^{(0)}}{v_g}}\dd\tau_0 e^{-i\tilde{\Omega}\tau_0}e_0(\va{X}^{(1)};t^{(1)})\notag\\
	&+\bigg(-\frac{\mathcal{V}^2}{v_g}\bigg)^2\sum_{\lambda_0=1,\lambda_1\neq\lambda_0}^{N-1}e^{i(\Omega/v_g)x_{\lambda_1}}\prod_{m=0}^{1}\Theta(x_{\lambda_m}^{(m)})\Theta(t^{(m+1)})\int_0^{\frac{x_{\lambda_0}^{(0)}}{v_g}}\dd\tau_0 e^{-i\tilde{\Omega}\tau_0}\mathcal{T}_{\lambda_1}(\va{X}^{(1)};t^{(1)}).\label{a72}
\end{align}
In deriving the above expansion, we have made use of Lemma 1 (c.f. Appendix C) to express $x_{\lambda_0}^{(0)}+x_{\lambda_1}^{(1)}=x_{\lambda_1}$. Further, in view of the fact that the $\lambda_0^{th}$ component of $\va{X}^{(1)}$ is $x_{\lambda_1}^{(1)}=v_g\tau_0-x_{\lambda_0}^{(0)}$, and $\tau_0<{x_{\lambda_0}^{(0)}}/{v_g}$ due to constraints on the integration range, the coordinate $x_{\lambda_0}^{(1)}<0$, and therefore, in writing down Eq. \ref{a72}, we have taken care to exclude the non-contributing terms with $\lambda_1=\lambda_0$.

Following this, we can now substitute the expression for $\mathcal{T}_{\lambda_1}(\va{X}^{(1)};t^{(1)})$, and expand the series further to next order in $\mathcal{V}^2/v_g$. A continued application of this generic mapping scheme finally leads us to the full expansion of $e(\va{x}_{N-1};t)$ expressed purely in terms of the known function $e_0$. Interestingly, the iteration terminates after a finite number of steps, since there exist only a finite number of excitation pathways. Th next iteration following Eq. \ref{a72} yields
\begin{align}
	e(\va{X}^{(0)};t^{(0)})=&e_0(\va{X}^{(0)};t^{(0)})-\frac{\mathcal{V}^2}{v_g}\sum_{\lambda_0=1}^{N-1}e^{i(\Omega/v_g)x_{\lambda_0}}\Theta(x_{\lambda_0}^{(0)})\Theta(t^{(1)})\int_0^{\frac{x_{\lambda_0}^{(0)}}{v_g}}\dd\tau_0 e^{-i\tilde{\Omega}\tau_0}e_0(\va{X}^{(1)};t^{(1)})\notag\\&+\bigg(-\frac{\mathcal{V}^2}{v_g}\bigg)^2\sum_{\{\lambda_0,\lambda_1\}\in\mathcal{P}_2}e^{i(\Omega/v_g)x_{\lambda_{1}}}\prod_{m=0}^{1}\Bigg[\Theta(x_{\lambda_m}^{(m)})\Theta(t^{(m+1)})\int_0^{\frac{x_{\lambda_m}^{(m)}}{v_g}}\dd{\tau_m}e^{-i\tilde{\Omega}\tau_m}\Bigg]e(\va{X}^{(2)};t^{(2)})\notag\\
	=&e_0(\va{X}^{(0)};t^{(0)})-\frac{\mathcal{V}^2}{v_g}\sum_{\lambda_0=1}^{N-1}e^{i(\Omega/v_g)x_{\lambda_0}}\Theta(x_{\lambda_0}^{(0)})\Theta(t^{(1)})\int_0^{\frac{x_{\lambda_0}^{(0)}}{v_g}}\dd\tau_0 e^{-i\tilde{\Omega}\tau_0}e_0(\va{X}^{(1)};t^{(1)})\notag\\&+\bigg(-\frac{\mathcal{V}^2}{v_g}\bigg)^2\sum_{\{\lambda_0,\lambda_1\}\in\mathcal{P}_2}e^{i(\Omega/v_g)x_{\lambda_{1}}}\prod_{m=0}^{1}\Bigg[\Theta(x_{\lambda_m}^{(m)})\Theta(t^{(m+1)})\int_0^{\frac{x_{\lambda_m}^{(m)}}{v_g}}\dd{\tau_m}e^{-i\tilde{\Omega}\tau_m}\Bigg]e_0(\va{X}^{(2)};t^{(2)})\notag\\
	&+\bigg(-\frac{\mathcal{V}^2}{v_g}\bigg)^3\sum_{\{\lambda_0,\lambda_1,\lambda_2\}\in\mathcal{P}_3}e^{i(\Omega/v_g)x_{\lambda_2}}\prod_{m=0}^{2}\Bigg[\Theta(x_{\lambda_m}^{(m)})\Theta(t^{(m+1)})\Bigg]\int_0^{\frac{x_{\lambda_1}^{(1)}}{v_g}}\dd\tau_1\int_0^{\frac{x_{\lambda_0}^{(0)}}{v_g}}\dd\tau_0 e^{-i\tilde{\Omega}(\tau_0+\tau_1)}\mathcal{T}_{\lambda_2}(\va{X}^{(2)};t^{(2)}),
\end{align}
Once again, Lemma 1 was invoked to write $x_{\lambda_0}^{(0)}+x_{\lambda_1}^{(1)}+x_{\lambda_2}^{(2)}=x_{\lambda_2}$, and terms with $\lambda_2=\lambda_1$ and $\lambda_2=\lambda_0$ were categorically dropped from the final sum in Eq. (74) since these would not contribute. This is engendered by the fact that both the coordinates $x_{\lambda_0}^{(2)}$ and $x_{\lambda_1}^{(2)}$ are forced to be negative. Further, in line with the algebraic prescription enunciated earlier, we have invoked the notations for the permutation sets $\{\mathcal{P}_j\}$ to indicate non-overlapping choices for the $\{\lambda_j\}$ variables. Clearly, by recurrently employing the combination of Eqs. \ref{a68} and \ref{a69}, we will keep obtaining increasingly higher powers of $\mathcal{V}^2/v_g$. After a total of $N-2$ iterations, this leads to
\begin{align}
	e(\va{X}^{(0)};t^{(0)})&=e_0(\va{X}^{(0)};t^{(0)})\notag\\
	&+\sum_{j=1}^{N-2}\bigg(-\frac{\mathcal{V}^2}{v_g}\bigg)^j\sum_{\mathbf{\Lambda}_j\in\mathcal{P}_j}e^{i(\Omega/v_g)x_{\lambda_{j-1}}}\prod_{m=0}^{j-1}\Bigg[\Theta(x_{\lambda_m}^{(m)})\Theta(t^{(m+1)})\int_0^{\frac{x_{\lambda_m}^{(m)}}{v_g}}\dd{\tau_m}e^{-i\tilde{\Omega}\tau_m}\Bigg]e_0(\va{X}^{(j)};t^{(j)})\notag\\
	&+\hspace{1mm}\bigg(-\frac{\mathcal{V}^2}{v_g}\bigg)^{N-1}\sum_{\mathbf{\Lambda}_{N-1}\in\mathcal{P}_{N-1}}e^{i(\Omega/v_g)x_{\lambda_{N-2}}}\prod_{m=0}^{N-2}\Bigg[\Theta(x_{\lambda_m}^{(m)})\Theta(t^{(m+1)})\Bigg]\prod_{m=0}^{N-3}\Bigg[\int_0^{\frac{x_{\lambda_m}^{(m)}}{v_g}}\dd{\tau_m}e^{-i\tilde{\Omega}\tau_m}\Bigg]\mathcal{T}_{\lambda_{N-2}}(\va{X}^{(N-2)};t^{(N-2)})\notag\\
	=&e_0(\va{X}^{(0)};t^{(0)})\notag\\
	&+\sum_{j=1}^{N-2}\bigg(-\frac{\mathcal{V}^2}{v_g}\bigg)^j\sum_{\mathbf{\Lambda}_j\in\mathcal{P}_j}e^{i(\Omega/v_g)x_{\lambda_{j-1}}}\prod_{m=0}^{j-1}\Bigg[\Theta(x_{\lambda_m}^{(m)})\Theta(t^{(m+1)})\int_0^{\frac{x_{\lambda_m}^{(m)}}{v_g}}\dd{\tau_m}e^{-i\tilde{\Omega}\tau_m}\Bigg]e_0(\va{X}^{(j)};t^{(j)})\notag\\
	+\hspace{1mm}&\bigg(-\frac{\mathcal{V}^2}{v_g}\bigg)^{N-1}\sum_{\mathbf{\Lambda}_{N-1}\in\mathcal{P}_{N-1}}e^{i(\Omega/v_g)x_{\lambda_{N-2}}}\prod_{m=0}^{N-2}\Bigg[\Theta(x_{\lambda_m}^{(m)})\Theta(t^{(m+1)})\int_0^{\frac{x_{\lambda_m}^{(m)}}{v_g}}\dd{\tau_m}e^{-i\tilde{\Omega}\tau_m}\Bigg]e(\va{X}^{(N-1)};t^{(N-1)}).\label{a74}
\end{align}
The final terms in Eq. \ref{a74}, which are proportional to $(\mathcal{V}^2/v_g)^{N-1}$, are non-zero only when $x_{\lambda_m}^{(m)}>0$ $\forall m\in\{0,1,2,...,N-2\}$, i.e., when $0<x_{\lambda_0}<x_{\lambda_1}<x_{\lambda_2}<...<x_{\lambda_{N-2}}$. However, since the integration variable $\tau_m$ is cut off at $x_{\lambda_m}^{(m)}$, we must have $v_g\tau_m-x_{\lambda_m}^{(m)}<0$ for each of the participating integer $m\in\{0,1,2,\hdots,N-2\}$. Consolidating these factors, we infer that
\begin{align}
	{x}_i^{(N-1)}=\underbrace{(x_i-x_{\lambda_{N-2}})}_{<0}+\sum_{l=0}^{N-2}\underbrace{(v_g\tau_l-x_{\lambda_l}^{(l)})}_{<0}\delta_{i,\lambda_l}<0,\hspace{2mm} \forall \hspace{1mm} i\in\{1,2,...,N-1\}.
\end{align}
This implies, from Eq. \ref{a69}, that 
\begin{align}
	e(\va{X}^{(N-1)};t^{(N-1)})=e_0(\va{X}^{(N-1)};t^{(N-1)}), 
\end{align}
and therefore, no further iterations would be necessary. The resulting expression reads
\begin{align}
			e(\va{x}_{N-1};t)=&\hspace{1mm}e_0(\va{x}_{N-1};t)\notag\\
			&+\sum_{j=1}^{N-1}\bigg(-\frac{\mathcal{V}^2}{v_g}\bigg)^j\sum_{\mathbf{\Lambda}_j\in\mathcal{P}_j}e^{i(\Omega/v_g)x_{\lambda_{j-1}}}\prod_{m=0}^{j-1}\Bigg[\Theta(x_{\lambda_m}^{(m)})\Theta(t^{(m+1)})\int_0^{\frac{x_{\lambda_m}^{(m)}}{v_g}}\dd{\tau_m}e^{-i\tilde{\Omega}\tau_m}\Bigg]e_0(\va{X}^{(j)};t^{(j)}),\label{a77}
\end{align}
wherein, we have now switched back to our original spacetime coordinates $\va{X}^{(0)}=\va{x}_{N-1}$ and $t^{(0)}=t$. It is worth remarking that on account of the coordinate-exchange symmetry embedded into the starting wave equations, no additional symmetrization was imperative in solving the latter and the derived wave function automatically reflects the correct permutation symmetry. While calculating the various higher order multivariate integrals might appear to be a daunting task at hand, leveraging the set-transformation lemmas stated in Appendix C can simplify these integrals into a simple product of identical single-variable integrals. This allows us to compute all arbitrary orders in the above expansion by cashing in on the already known expression for $e_0$. The definition laid out in Eq. \ref{e17} translates to
\begin{align}
	e_0(\va{X}^{(j)};t^{(j)})&=-\frac{\sqrt{N}i\mathcal{V}}{(2\pi)^{N/2}(\omega_k-\tilde{\Omega})}\bigg(e^{ik\sum_{l=1}^{N-1}x_l^{(j)}}\bigg)\bigg(e^{-i(N\omega_k-\Omega)t^{(j)}}\bigg)\bigg(e^{i(\omega_k-\tilde{\Omega})t^{(j)}}-1\bigg)\notag\\
	&=-\frac{\sqrt{N}i\mathcal{V}}{(2\pi)^{N/2}(\omega_k-\tilde{\Omega})}\bigg(e^{ik\sum_{l=1}^{N-1}x_l}\bigg)\bigg(e^{-iNkx_{\lambda_{j-1}}}\bigg)\bigg(e^{i\omega_k\sum_{l=0}^{j-1}\tau_l}\bigg)\bigg(e^{-i(N\omega_k-\Omega)t^{(j)}}\bigg)\bigg(e^{i(\omega_k-\tilde{\Omega})t^{(j)}}-1\bigg),\label{a78}
\end{align}
where Lemma 2 has been invoked to simplify the coordinates $\va{X}^{(j)}=\{x_i^{(j)}\}$ in terms of $\va{X}^{(0)}=\{x_i\}$. Now since 
\begin{align}
	\int_{0}^{{x_{\lambda_m}^{(m)}}/{v_g}}\dd{\tau_m}e^{i(\omega_k-\tilde{\Omega})\tau_m}=-i[e^{i(\omega_k-\tilde{\Omega})x_{\lambda_m}^{(m)}/v_g}-1]/(\omega_k-\tilde{\Omega}),
\end{align}
the term proportional to $\bigg(-\dfrac{\mathcal{V}^2}{v_g}\bigg)^j$ in Eq. \ref{a78}, which involves a product of $j$ integrals over the \textit{independent} variables $\{\tau_0,\tau_1,\hdots,\tau_{j-1}\}$, can be straightforwardly evaluated.  This consideration leads directly to Eq. \ref{e18} and formally completes our derivation of the time-varying wave function underpinning the atomic excitation dynamics.

\section{Appendix E: Scattering dynamics of a Gaussian wave packet} \label{se}
Here we consider our incident radiation field to be modeled by a normalized, multiphoton, Gaussian wavepacket centered around the wavenumber $k$ and comprising of a finite number of uncorrelated photons. In real space, it would be represented by the initial conditions
\begin{align}
	f(\va{x}_N;0)&=\bigg(\frac{1}{(\pi d)^{N/4}}e^{-\frac{1}{2d^2}\sum_{j=1}^Nx_j^2}\bigg)\hspace{1mm}e^{ik\sum_{j=1}^Nx_j},\notag\\
	e(\va{x}_{N-1};0)&=0.
\end{align}
If the parameter $d$, which has the dimension of length, is taken to be large enough, this would approximate a plane wave. With this new parametrization, the wave function $e_0(\va{x}_{N-1};t)$ can be re-expressed as
\begin{align}
	e_0(\va{x}_{N-1};t)&=\sqrt{N}\mathcal{V}e^{-i(N\omega_k-\Omega) t}\bigg(e^{i\sum_{j=1}^{N-1}(k-iv_gt/d^2)x_j}\bigg)\bigg(e^{-\frac{1}{2d^2}\sum_{j=1}^{N-1}x_j^2}\bigg)\bigg(e^{-v_g^2t^2/d^2}\bigg)\mathcal{I}_0(t),
\end{align}
where the function $\mathcal{I}_0(t)$ can be simplified in terms of error functions,
\begin{align}
	\mathcal{I}_0(t)&=\int_{0}^{t}\dd\tau \hspace{1mm} e^{-v_g^2\tau^2/d^2}e^{i(\omega_k-\tilde{\Omega}-iv_g^2t/d^2)\tau}\notag\\
	&=\frac{1}{2}\sqrt{\frac{\pi}{a}}\hspace{1mm}e^{b_0^2/(4a_0)}\bigg[\erf{\sqrt{a_0}\bigg(t+\frac{b_0}{2a_0}}\bigg)-\erf\bigg({\frac{b_0}{2\sqrt{a_0}}}\bigg)\bigg],
\end{align}
with the constants $a_0={v_g^2}/{(2d^2)}$, and $b_0=-i(\omega_k-\tilde{\Omega}-iv_g^2t/d^2)$. Similarly, we can calculate
\begin{align}
	e_0(\va{X}^{(j)};t^{(j)})&=\sqrt{N}\mathcal{V}e^{-i(N\omega_k-\Omega) t^{(j)}}\bigg(e^{i\sum_{l=1}^{N-1}(k-iv_gt^{(j)}/d^2)x_l^{(j)}}\bigg)\bigg(e^{-\frac{1}{2d^2}\sum_{l=1}^{N-1}(x_l^{(j)})^2}\bigg)\bigg(e^{-v_g^2(t^{(j)})^2/d^2}\bigg)\mathcal{I}_0(t^{(j)}).\label{a83}
\end{align}
We, therefore, need to express the sums $\sum_{l=1}^{N-1}x_l^{(j)}$ and $\sum_{l=1}^{N-1}(x_l^{(j)})^2$ in terms of the original spacetime variables $(x_1,x_2,x_3,\hdots,x_{N-1};t)$. The corresponding reduced expressions are presented below for reference:
\begin{align}
	\sum_{l=1}^{N-1}x_l^{(j)}&=\bigg(\sum_{l=1}^{N-1}x_l\bigg)-Nx_{\lambda_{j-1}}+v_g\sum_{m=0}^{j-1}\tau_m,\notag\\
	\sum_{l=1}^{N-1}(x_l^{(j)})^2&=\bigg(\sum_{l=1}^{N-1}x_l^2\bigg)-2x_{\lambda_{j-1}}\bigg(\sum_{l=1}^{N-1}x_l\bigg)+Nx_{\lambda_{j-1}}^2+\bigg[v_g^2\sum_{m=0}^{j-1}\tau_{m}^2+2v_g\sum_{m=0}^{j-1}\bigg(x_{\lambda_{m-1}}+x_{\lambda_{j-1}}\bigg)\tau_{m}\bigg].
\end{align}
With these simplifications, we obtain the solution to the Gaussian case by evaluating Eq. \ref{a83}, which is given by\\
\fbox{
	\addtolength{\linewidth}{-2\fboxsep}%
	\addtolength{\linewidth}{-2\fboxrule}%
	\begin{minipage}{\linewidth}
		\begin{align}
			e(\va{x}_{N-1};t)=&\frac{\sqrt{N}\mathcal{V}}{(\pi d)^{N/4}}e^{-i(\omega_k-\Omega+iv_gx_j/d^2) t}\bigg(\prod_{l=1}^{N-1}e^{i(kx_l-\omega_kt)}\bigg)\bigg(e^{-\frac{1}{2d^2}\sum_{j=1}^{N-1}x_j^2}\bigg)\bigg(e^{-v_g^2t^2/d^2}\bigg)\Bigg[\mathcal{I}_0(t)\notag\\&+\sum_{j=1}^{N-1}\bigg(-\frac{\mathcal{V}^2}{v_g}\bigg)^j\Theta(t^{(j)})\sum_{\mathbf{\Lambda}_j\in\mathcal{P}_j}\mathcal{I}_0(t^{(j)})\exp{-\frac{(N-2)v_g^2[t^{(j)2})-t^2]}{2d^2}}\prod_{m=0}^{j-1}\mathcal{I}_m\bigg(\frac{x_{\lambda_m}^{(m)}}{v_g}\bigg)\Theta(x_{\lambda_m}^{(m)})\hspace{1mm}\Bigg]\hspace{1mm},
		\end{align}
	\end{minipage}\\
}

in which we have introduced the more general functions 
\begin{align}
	\mathcal{I}_m(t)&=\int_{0}^{t}\hspace{1mm}\dd\tau \hspace{1mm} e^{-v_g^2\tau^2/d^2}e^{i(\omega_k-\tilde{\Omega}-iv_g^2t^{(m)}/d^2)\tau}\notag\\
	&=\frac{1}{2}\sqrt{\frac{\pi}{a_m}}e^{b_m^2/(4a_m)}\hspace{1mm}\bigg[\erf{\sqrt{a_m}\bigg(t+\frac{b_m}{2a_m}}\bigg)-\erf\bigg({\frac{b_m}{2\sqrt{a_m}}}\bigg)\bigg],
\end{align}
with $a_m=a_0={v_g^2}/{(2d^2)}$, and $b_m=-i(\omega_k-\tilde{\Omega}-iv_g^2t^{(m)}/d^2)$.

\section{Appendix F: Relevant integrals for demonstrating Rabi oscillation} \label{sf}
We hereby compute the probability of atomic excitation, following the expression in Eq. \ref{e35}. Since we focus on the quasimonochromatic case, we can approximate $(v_g\kappa)t\ll1$ for the timescale of interest. In this limit, the net probability can be computed as a sum of the following four terms:
\begin{align}
	p_0&=\abs{\mathcal{A}_{\kappa}}^2{\abs{\phi_{\kappa}(t)}^2}\bigg(\int\dd{x}e^{-2\kappa\abs{x}}\bigg)^{N-1},\label{a87}\\
	p_1&\approx\abs{\mathcal{A}_{\kappa}}^2{\phi_{\kappa}^*(t)}\sum_{j=1}^{N-1}(1-\mathbf{t}_k)^j\frac{1}{\kappa^{N-j-1}}\notag\sum_{\mathbf{\Lambda}_j\in\mathcal{P}_j}\prod_{m=0}^{j-1}\bigg[\int\dd{{x}_{\lambda_m}}\phi_{\kappa}\bigg(\frac{x_{\lambda_m}^{(m)}}{v_g}\bigg)\Theta(x_{\lambda_m}^{(m)})\bigg]\phi_{\kappa}(t^{(j)})\Theta(t^{(j)}),\\
	p_2&=p_1^*,\label{a88}\\
	p_3&\approx\abs{\mathcal{A}_{\kappa}}^2\sum_{i,j=1}^{N-1}(1-\mathbf{t}_k^*)^i\hspace{1mm}(1-\mathbf{t}_k)^j\sum_{\mathbf{\Lambda}_i\in\mathcal{P}_i,\mathbf{\Lambda'}_j\in\mathcal{P}_j}\frac{1}{\kappa^{N-n_{ij}-1}}\cross\notag\\&\int\prod_{m=0}^{i-1}\prod_{n=0}^{j-1}\Bigg[\phi_{\kappa}^*\bigg(\frac{x_{\lambda_m}^{(m)}}{v_g}\bigg)\phi_{\kappa}\bigg(\frac{x_{\lambda'_n}^{(n)}}{v_g}\bigg)\Theta(x_{\lambda_m}^{(m)})\Theta(x_{\lambda'_n}^{(n)})\Bigg]\phi_{\kappa}^*(t^{(i)})\phi_{\kappa}(t^{'(j)})\Theta(t^{(i)})\Theta(t^{'(j)}).\label{a89}
\end{align}
The quantity $n_{ij}=\abs{\mathbf{\Lambda}_i\cup\mathbf{\Lambda'}_j}$ denotes the number of elements in the union between the two sets $\mathbf{\Lambda}_i$ and $\mathbf{\Lambda'}_j$, and it can vary from $\max\{i,j\}$ to $i+j$. In writing the final defining integral for $p_3$, we used the compact notation $t^{'(j)}=t-\dfrac{x_{\lambda'_{j-1}}}{v_g}$. The first integral in Eq. \ref{a87} can be promptly calculated, leading to 
\begin{align}
	p_0&\approx\frac{{N\mathcal{V}^2\kappa}\hspace{1mm}\abs{\phi_{\kappa}(t)}^2}{(\omega_k-\Omega)^2+(\Gamma-v_g\kappa)^2}\notag\\
	&=\frac{2{{N}(v_g\kappa)\hspace{1mm}\Gamma}\hspace{1mm}[1+e^{-2(\Gamma-v_g\kappa) t}-2e^{-(\Gamma-v_g\kappa)t}\cos(\omega_k-\Omega)t]}{(\omega_k-\Omega)^2+(\Gamma-v_g\kappa)^2},
\end{align}
where we have set $e^{-2v_g\kappa t}\approx 1$. Since $p_0$ scales linearly with the number of photons, it merely betokens the part of the full excitation probability applicable in the linear optical regime without any interference stemming from photon-photon correlations. All information regarding photon-photon correlations is fully encoded into the contributions from $p_2$, $p_3$, and $p_4$. The orders of the exponent of $\kappa$ in these expressions (c. f. Eqs. \ref{a88},\ref{a89}) entails from two salient factors. For instance, let us reflect on the case of Eq. \ref{a88}. If $x_l$ is a coordinate such that $l\notin\mathbf{\Lambda}_{j}=(\lambda_0,\lambda_1,\lambda_2,\hdots,\lambda_{j-1})$, the integrand associated with this particular coordinate can be integrated out easily, yielding
\begin{align}
	\int_{-\infty}^{\infty}\dd{x}_le^{-2\kappa\abs{x_{l}-v_gt}}=\frac{1}{\kappa}.\label{a91}
\end{align}
Since there are $N-j-1$ such coordinates, this spews out an overall factor of $\dfrac{1}{\kappa^{N-j-1}}$. For the rest of the $j$ coordinates, i.e., $\{x_{\lambda_0},x_{\lambda_1},x_{\lambda_2,\hdots,x_{\lambda_{j-1}}}\}$, the integration is far more involved but restricted over a finite interval, courtesy of the participation from the Heaviside-theta functions. To be precise, the $j^{th}$-order term (which is proportional to $(1-\mathbf{t}_k)^j$) appearing in \ref{a88} pertains to a curtailed integration range due to the constraint $0<x_{\lambda_0}<x_{\lambda_1}<x_{\lambda_2}<\hdots<x_{\lambda_{j-1}}<v_gt$. Thus, the overall integral ensuing from these coordinates can be expressed as
\begin{align}
	\sum_{\mathbf{\Lambda}_j\in\mathcal{P}_j}\int_0^{v_gt}\dd{{x}_{\lambda_{j-1}}}\int_0^{{x}_{\lambda_{j-1}}}\dd{{x}_{\lambda_{j-2}}}\hdots\int_0^{{x}_{\lambda_{2}}}\dd{{x}_{\lambda_{1}}}\int_0^{{x}_{\lambda_{1}}}\dd{{x}_{\lambda_{0}}}\hspace{1mm}e^{-2\kappa\sum_{l\in\mathbf{\Lambda}_{j}}\abs{x_l-v_gt}}\hspace{1mm}\bigg[\prod_{m=0}^{j-1}\phi_{\kappa}\bigg(\frac{x_{\lambda_m}^{(m)}}{v_g}\bigg)\bigg]\hspace{1mm}\phi_{\kappa}(t^{(j)}).
\end{align}
Clearly, the integration ranges for these coordinates are finite and the residual integrand is also bounded above. In particular, since $(v_g\kappa)t\ll 1$, this allows us to make the following approximations inside the above integral: $e^{-2\kappa\abs{x_{\lambda_m}-v_gt}}\approx 1$ and $\phi(z)=e^{i(\omega_k-\Omega)z}e^{-(\Gamma-v_g\kappa) z}-1\approx e^{i(\omega_k-\Omega)z}e^{-\Gamma z}-1$. This approximation scheme for arbitrarily small $\kappa$ obviously fails for improper integrals extending over an infinite range, as evidenced by Eq. \ref{a91}, since $\kappa$, being positive, serves as a regularizing parameter purging any risk of divergence. A similar argument could be invoked to justify the expression shown in Eq. \ref{a89}. Note that for any two selected sets $\mathbf{\Lambda}_i$ and $\mathbf{\Lambda'}_j$ in this equation, the total number of coordinates that do not belong to either one of these sets is equal to $N-n_{ij}-1$, which when integrated over, furnishes a factor of $\dfrac{1}{\kappa^{N-n_{ij}-1}}$. 

We now calculate the asymptotic expressions for $p_1$, $p_2$, and $p_3$. Since all of the integration coordinates in Eqs. \ref{a88},\ref{a89} are bounded above by $v_gt$, this implies that for $m\in\{0,1,2,\hdots,j-1\}$ we can proceed with the approximations 
\begin{align}
	\phi(x_{\lambda_m}^{(m)}/v_g)&\approx -(\Gamma/v_g)x_{\lambda_m}^{(m)},\notag\\
	\phi(t^{(j)})&\approx-\Gamma t^{(j)}.
\end{align}
Further, on resonance, we have $\mathbf{t}_k=-1$. Now considering that there are $\Perm{N-1}{j}$ possible ways of selecting $j$ coordinates and each of these permutations contributes equally to the integration, the expression for $p_1$ can be simplified as
\begin{align}
	p_1&\approx\sum_{j=1}^{N-1}\Perm{N-1}{j}\hspace{1mm}(-1)^j\hspace{1mm}(2\kappa \Gamma)^j\hspace{1mm}\bigg(\frac{1}{v_g}\bigg)^{j+1}\hspace{1mm}\bigg[2N(v_g\kappa)\Gamma t^2\bigg]\hspace{1mm}\mathcal{K}_j(v_gt),\label{a94}
\end{align}
wherein we have introduced a class of functions defined recursively through
\begin{align}
	\mathcal{K}_j(x)=\int_0^x\dd{x'} (x-x')\hspace{1mm}\mathcal{K}_{j-1}(x'),
\end{align}
with the known boundary term $\mathcal{K}_0(x)=x$. The above equation is simply a shorthand notation for the multivariate integral
\begin{align}
	\mathcal{K}_j(x)=\int_0^{x}\dd{x}_j(x-x_j)\int_0^{x_j}\dd{x}_{j-1}(x_j-x_{j-1})\hdots\int_{0}^{x_3}\dd{x}_2(x_3-x_2)\int_0^{x_2}\dd{x}_1(x_2-x_1)\hspace{1mm}x_1,
\end{align}
which satisfies the differential recurrence relation
\begin{align}
	\frac{\dd^2\mathcal{K}_j(x)}{\dd{x}^2}=\mathcal{K}_{j-1}(x).
\end{align}
Exploiting the above relation iteratively, supplemented by the condition $\mathcal{K}_0(x)=x$, we can prove the following:
\begin{align}
	\mathcal{K}_1(x)&=\frac{x^{3}}{3!},\notag\\
	\mathcal{K}_2(x)&=\frac{x^{5}}{5!},\notag\\
	\mathcal{K}_3(x)&=\frac{x^{7}}{7!},\notag\\
	&\vdots\notag\\
	\mathcal{K}_j(x)&=\frac{x^{2j+1}}{(2j+1)!}.
\end{align}
Consequently, substituting for this into Eq. \ref{a94}, we get the simplified form
\begin{align}
	p_1&\approx\sum_{j=1}^{N-1}\bigg[\frac{\Perm{N-1}{j}}{N^j (2j+1)!}\bigg]\hspace{1mm}(-1)^j\hspace{1mm}\bigg[2N(v_g\kappa)\Gamma t^2\bigg]^{j+1}.
\end{align}

It is at this moment that we can afford to make an asymptotic analysis of this expression. Since we are in the strong-pumping regime, we can make appropriate reductions by taking the limit $N\rightarrow\infty$. Particularly, we can approximate $\Perm{N-1}{j}\approx\Perm{N}{j}=N(N-1)(N-2)\hdots(N-j-2)(N-j-1)\approx N^j$ for any finite $j$, and this helps us reduce the foregoing into an infinite series,
\begin{align}
	p_1&\approx\sum_{j=1}^{\infty}(-1)^j\hspace{1mm}\frac{1}{(2j+1)!}\hspace{1mm}(g^2t^2)^{j+1}.
\end{align}
The calculation of $p_3$ as constructed in Eq. \ref{a89} requires more subtle considerations. In particular, only the terms with $n_{ij}=i+j$ are important enough to be computed as the rest of the integrals will be higher order in $\Gamma t$, and therefore, can contribute very little when the spontaneous relaxation rate $\Gamma$ remains strongly subordinated in comparison to the multiphoton Rabi frequency $g=\sqrt{2N(v_g\kappa)\Gamma}$. In other words, if the sets $\mathbf{\Lambda}_i$ and $\mathbf{\Lambda'}_j$ in Eq. \ref{a89} contain any overlapping elements, the ensuing integrals will remain negligible. Neglecting the subleading contribution from these integrals dramatically simplifies the analysis. It is, however, to be borne in mind that this argument only holds potential in the large-$N$ limit as the pumping rate is ramped up prodigiously thereby creating an appreciable Rabi frequency. Consequently, while the effect of the Rabi frequency $g$ can always be factored in, the competing influence of spontaneous emission from the atom can be swept under the rug. The term corresponding to $n_{ij}=i+j$  is the only term which can be reduced principally in terms of the Rabi frequency, while the rest of the terms underpin the smothering influence of spontaneous emission on the atomic transition (which we demonstrate below). First, let us comment on the case when $n_{ij}< i+j$. Denoting the corresponding contribution as $p_{n_{ij}}'$, we find the following bound on its magnitude:
\begin{align}
	\max\{p_{n_{ij}}'\}&\sim\sum_{i=1}^N\sum_{j=1}^N \hspace{1mm}\frac{(2N(v_g\kappa)\Gamma t^2)^{n_{ij}+1}}{{n_{ij}!}}\hspace{1mm}\mathcal{O}(\Gamma t)^{i+j-{n_{ij}}}.
\end{align}
This is certainly not a tight bound, and in fact, it is not trivial to obtain a tight bound on $p_{n_{ij}}'$. It is an order-of-magnitude estimate of its algebraic maximum. However, it serves the desired purpose demonstrating that its contribution to the full probability must be subleading in $\Gamma t$ unless $n_{ij}=i+j$. Whenever $\max\{i,j\}<n_{ij}< i+j$, the above expression would contain terms which are at least linear order in $\Gamma t$, aside from integer powers of $gt$. However, the choice $n_{ij}=i+j$ would ensure that the corresponding contribution only contains powers of $g^2t^2$. In that case, the asymptotic ($N\rightarrow\infty$) limit of $p_3$ can be conveniently expressed entirely in terms of the contribution due to $n_{ij}=i+j$. This subtle consideration begets the following approximate form for $p_3$:
\begin{align}
	p_3&\approx\sum_{i=1}^{N-1}\sum_{j=1}^{N-1}p_{i+j}'\notag\\
	&=\frac{2N(v_g\kappa)\Gamma}{\Gamma^2}\sum_{i=1}^{N-1}\sum_{j=1}^{N-1}\bigg(\Perm{N-1}{i}\Perm{N-1-i}{j}\bigg)\hspace{1mm}(2\kappa)^{i+j}\hspace{1mm}\bigg(-\frac{\Gamma}{v_g}\bigg)^{i+j+2}\mathcal{K}_i(v_gt)\hspace{1mm}\mathcal{K}_j(v_gt)\notag\\
	&=\sum_{i=1}^{N-1}\sum_{j=1}^{N-1} \bigg[\frac{\Perm{N-1}{i}\hspace{1mm}\Perm{N-1-i}{j}}{N^{i+j}}\bigg]\hspace{1mm}\bigg[\frac{(-1)^{i+j}}{(2i+1)!\hspace{1mm}(2j+1)!}\bigg]\hspace{1mm}\bigg[2N(v_g\kappa)\Gamma t^2\bigg]^{i+j+1}\notag\\
	&\approx \sum_{i=1}^N\sum_{j=1}^N \hspace{1mm}\frac{(-1)^{i+j}}{(2i+1)!\hspace{1mm}(2j+1)!}\hspace{1mm}(2N(v_g\kappa)\Gamma t^2)^{i+j+1}\notag\\
	&\approx\sum_{n=2}^{\infty}(-1)^n\hspace{1mm}\bigg[\sum_{j=1}^{n-1}\frac{1}{(2j+1)!\hspace{1mm}(2(n-j)+1)!}\bigg]\hspace{1mm}(g^2 t^2)^{n+1}.
\end{align}
In the penultimate step, we assumed $\Perm{N-1}{i}\hspace{1mm}\Perm{N-1-i}{j}=\dfrac{(N-1)!}{(N-1-i)!}.\dfrac{(N-1-i)!}{(N-1-i-j)!}=\Perm{N-1}{i+j}\approx N^{i+j}$. Thereupon, summing up the various contributions, we find the total probability in the asymptotic large-field limit to be given by
\begin{align}
	p_N^{(e)}(t)&=p_0+p_1+p_2+p_3\notag\\
	&\approx g^2t^2+\sum_{n=1}^{\infty}(-1)^n\hspace{1mm}\frac{2}{(2n+1)!}\hspace{1mm}(g^2t^2)^{n+1}+\sum_{n=2}^{\infty}(-1)^n\bigg[\sum_{j=1}^{n-1}\frac{1}{(2j+1)!\hspace{1mm}(2(n-j)+1)!}\bigg]\hspace{1mm}(g^2t^2)^{n+1}\notag\\
	&=g^2t^2-\frac{1}{3}g^4t^4+\sum_{n=2}^{\infty}(-1)^n\hspace{1mm}\chi_n\hspace{1mm}(g^2t^2)^{n+1},\label{a103}
\end{align}
wherein the coefficients $\chi_n$ feature the series expansions
\begin{align}
	\chi_n&=\sum_{j=0}^{n}\frac{1}{(2j+1)!\hspace{1mm}(2(n-j)+1)!}\notag\\
	&=\frac{1}{(2n+2)!}\sum_{j\rightarrow\text{odd}}^{2n+2} \Comb{2n+2}{j}.
\end{align}
This series can be exactly computed by invoking the following property of a binomial series:
\begin{align}
	\sum_{j\rightarrow\text{odd}}^n \Comb{n}{j}\hspace{1mm}x^j=\frac{(1+x)^n-(1-x)^n}{2}.
\end{align}
In particular, setting $x=1$ and replacing $n$ by $2n+2$ yields the desired coefficients,
\begin{align}
	\chi_n=\frac{2^{2n+1}}{(2n+2)!}.
\end{align}
Plugging this result into Eq. \ref{a103} leads to our final expression for the excitation probability,
\begin{align}
	p_N^{(e)}(t)&\approx  g^2t^2-\frac{1}{3}g^4t^4+\frac{1}{2}\sum_{n=2}^{\infty}(-1)^n\frac{2^{2n+2}}{(2n+2)!}(g^2t^2)^{n+1}\notag\\
	&=\frac{1}{2}\sum_{n=1}^{\infty}(-1)^{n+1}\frac{(2gt)^{2n}}{(2n)!}\notag\\
	&=\frac{1}{2}(1-\cos 2gt)\notag\\
	&=\sin^2 gt.
\end{align}

\section{Appendix G: Quantum analogue to the semiclassical Rabi frequency} \label{sg}
In order to strike a clear correspondence with the semiclassical coupling parameter $g=\dfrac{\va{p}\vdot\va{\mathcal{E}}}{\hbar}$, let us quantify the characteristics of the electric-field strength as relevant to the second quantized (fully quantum optical) framework. Using the electric-field mode expansion in $k$-space, $E(x)=\sum_{k}\sqrt{\dfrac{\hbar\omega_k}{2\epsilon_0AL}}c_k^{\dagger}e^{-ikx}+\text{h.c.}$ \footnote{The polarization degree of freedom has been suppressed since the model apparatus involves an SPSM waveguide.}, the time-varying electric field envelope in real space can be obtained through Fourier transform, where $A$ is the cross-section of the pulse and $L$ is the waveguide quantization length:
\begin{align}
	E(x)&=\sum_{k}\sqrt{\frac{\hbar\omega_k}{2\epsilon_0AL}}c_k^{\dagger}e^{-ikx}+\text{h.c.}\notag\\
	&\approx\sqrt{\frac{\hbar\Omega}{2\epsilon_0AL}}\sqrt{\frac{L}{2\pi}}\int_{-\infty}^{\infty}\dd{k}\hspace{1mm}c^{\dagger}(k)e^{-ikx}+\text{h.c.}\notag\\
	&=\frac{1}{{2\pi}}\sqrt{\frac{\hbar\Omega}{2\epsilon_0A}}\int_{-\infty}^{\infty}\dd{k}\hspace{1mm}\int_{-\infty}^{\infty}\dd{x'}c^{\dagger}(x')e^{ik(x'-x)}+\text{h.c.}\notag\\
	&=\frac{1}{{2\pi}}\sqrt{\frac{\hbar\Omega}{2\epsilon_0A}}\int_{-\infty}^{\infty}\dd{x'}c^{\dagger}(x')\int_{-\infty}^{\infty}\dd{k}\hspace{1mm}e^{ik(x'-x)}+\text{h.c.}\notag\\
	&=\sqrt{\frac{\hbar\Omega}{2\epsilon_0A}}\int_{-\infty}^{\infty}\dd{x'}c^{\dagger}(x')\delta(x'-x)+\text{h.c.}\notag\\
	&=\sqrt{\frac{\hbar\Omega}{2\epsilon_0A}}c^{\dagger}(x)+\text{h.c.}
\end{align}
Here, in the second step, we have invoked the slowly-varying-envelope-approximation to approximate the frequency-dependent amplitude $\sqrt{\dfrac{\hbar\omega_k}{2\epsilon_0AL}}$ as $\sqrt{\dfrac{\hbar\Omega}{2\epsilon_0AL}}$, since the amplitude varies rather sluggishly around the resonance frequency $\omega_k=\Omega$ as compared to the phase factors $e^{\pm ikx}$. We have also transformed the discrete-mode creation operator $c_k^{\dagger}$ to its continuous-mode representation $c^{\dagger}(k)$, bearing in mind the sanctity of the bosonic commutation relation. Since $[c_k,c_{k'}^{\dagger}]=\delta_{k,k'}$ (a dimensionless identity) is required to transform into $[c(k),c^{\dagger}(k')]=\delta(k-k')$ (a dimensionful identity), we have employed the dimensionally scaled mapping $c_k^{\dagger}\rightarrow \sqrt{\dfrac{L}{2\pi}}c^{\dagger}(k)$ to preserve the commutation relation. This implicit scaling factor follows from the transformation property $\dfrac{L}{2\pi}\delta_{k,k'}\rightarrow \delta(k-k')$ \footnote{To prove this, it is enough to consider the discrete sum $\sum_{k'}\bigg(\dfrac{L}{2\pi}\bigg)\delta_{k,k'}=\dfrac{L}{2\pi}$, which, in the continuum limit of the guided mode frequencies, can also be expressed as $\int_{-\infty}^{\infty}\bigg(\dfrac{L\dd{k}}{2\pi}\bigg)\hspace{0.5mm}\delta(k-k')$, where we recognize $\dfrac{L}{2\pi}$ as representing the density of quantum states in $k$-space.}. Interestingly, the transition to the continuum limit has dispensed with the $L$-dependence, although the $A$-dependence persists. However, the dependence on pulse cross section is also propitiously entrenched in the expression for the atom-photon coupling strength $\mathcal{V}=\sqrt{2v_g\Gamma}=\dfrac{\abs{\va{\mathbf{{p}}}}}{\hbar}\sqrt{\dfrac{\hbar\Omega}{2\epsilon_0A}}$, where, to remind the reader, $\abs{\va{\mathbf{{p}}}}$ stands for the magnitude of the atomic dipole moment. Given the real-space representation of the electric field operator, the time-varying envelope of the electric field can be obtained as 
\begin{align}
	\mathcal{E}(t)&=\sqrt{\expval{E^-(0^-)E^+(0^-)}}_{\ket{\psi_N(t)}}\notag\\
	&=\sqrt{\frac{\hbar\Omega}{2\epsilon_0A}\bra{\psi_N(t)}c^{\dagger}(0^-)c(0^-)\ket{\psi_N(t)}},
\end{align}

In the limit of strong excitation, we can ignore the atomic effect on the pulse, in which case, we can conclude to a good approximation that
\begin{align}
	\mathcal{E}^2(t)&=\frac{N\hbar\Omega}{2\epsilon_0A}\int\dd{\va{x}_{N-1}}\abs{f(x_1,x_2,x_3,\hdots,x_{N-1},0^-;t)}^2+\frac{(N-1)\hbar\Omega}{2\epsilon_0A}\int\dd{\va{x}_{N-2}}\abs{e(x_1,x_2,x_3,\hdots,x_{N-2},0^-;t)}^2\notag\\
	&\approx\frac{N\hbar\Omega}{2\epsilon_0A}\int\dd{\va{x}_{N-1}}\abs{f(x_1,x_2,x_3,\hdots,x_{N-1},0^-;t)}^2.
\end{align}
Using the lowest-order approximation for $f$ from Eq. (13), we obtain 
\begin{align}
	\mathcal{E}^2(t)&\approx\frac{N\hbar\Omega}{2\epsilon_0A}\kappa^N\int\dd{\va{x}_{N-1}}e^{-2\kappa\sum_{j=1}^{N-1}\abs{x_j-v_gt}}\notag\\
	&=\frac{N\hbar\Omega\kappa}{2\epsilon_0A}e^{-2v_g\kappa t}.\notag\\
	\implies\mathcal{E}(t)&=\sqrt{\frac{N\hbar\Omega\kappa}{2\epsilon_0A}}e^{-v_g\kappa t}.
\end{align}
This implies that the electric field envelope incurs exponential attenuation in time at a rate of $1/(v_g\kappa)$ per unit time. With this pulse envelope function, the appropriate pulse area in the limit of $(v_g\kappa)t\ll 1$ (as appropriate to quasi-plane waves) is calculated to be
\begin{align}
	\frac{\abs{\va{\mathbf{{p}}}}}{\hbar}\int_0^t\mathcal{E}(\tau)\dd{\tau}&\approx\frac{\abs{\va{\mathbf{{p}}}}}{\hbar}\sqrt{\frac{N\hbar\Omega\kappa}{2\epsilon_0A}}t\notag\\
	&=\sqrt{2N(v_g\kappa)\Gamma} t,
\end{align}
which is precisely the argument of the sine function in Eq. \ref{e47}. Thus, the quantum-mechanically calculated pulse area is consistent with the area obtained for a semiclassical pulse, and therefore, the corresponding Rabi frequencies are conformable, i.e., $\Omega_R^{(\text{quantum})}=\Omega_R^{(\text{semiclassical})}$.

\end{document}